\RequirePackage[2024-06-01]{latexrelease}
\documentclass[twocolumn,breaklinks=true]{aastex701}

\usepackage[table]{xcolor}
\usepackage{graphicx}
\usepackage{amsmath}
\usepackage{amssymb}
\usepackage{tabularx}
\usepackage{color, colortbl}
\usepackage{hhline}
\usepackage{multirow}
\usepackage{makecell}
\usepackage{booktabs}
\usepackage{threeparttable}
\usepackage{array}
\usepackage{acronym}
\usepackage{xspace}
\newcolumntype{P}[1]{>{\centering\arraybackslash}p{#1}}

\acrodef{BPS}{binary population synthesis}
\acrodef{SN}{supernova}
\acrodef{PISN}{pair-instability supernova}
\acrodef{ECSN}{electron-capture supernova}
\acrodef{GW}{gravitational wave}
\acrodef{BBH}{binary black hole}
\acrodef{CE}{common envelope}
\acrodef{SMT}{stable mass transfer}
\acrodef{DCO}{double compact object}
\acrodef{ZAMS}{zero-age main sequence}
\acrodef{RLOF}{Roche-lobe overflow}
\acrodef{HR}{Hertzprung-Russell}

\def\CosmicMRRPerc{\ensuremath{\sim 45\%}\xspace}
\def\PosydonMRRPerc{\ensuremath{\sim 62\%}\xspace}
\def\MetisseMRRPerc{\ensuremath{\sim 63\%}\xspace}

\def\ThreeCodeOverlap{\ensuremath{1, 4, \text{and \xspace} 0} \xspace}

\def\CosmicMetisseOverlapN{\ensuremath{982}\xspace}
\def\CosmicPosydonOverlapN{\ensuremath{26}\xspace}
\def\MetissePosydonOverlapN{\ensuremath{27}\xspace}

\def\CosmicMetisseDSC{\ensuremath{0.137}\xspace}
\def\CosmicMetisseOverlapPerc{\ensuremath{13.7\%}\xspace}
\def\CosmicPosydonOverlapPerc{\ensuremath{0.3\%}\xspace}
\def\MetissePosydonOverlapPerc{\ensuremath{0.5\%}\xspace}

\def\CosmicBBHMergers{\ensuremath{8,975}\xspace}
\def\MetisseBBHMergers{\ensuremath{5,385}\xspace}
\def\PosydonBBHMergers{\ensuremath{6,369}\xspace}

\newcommand{\UChicago}{\affiliation{Department of Physics, The University of Chicago, 5640 South Ellis Avenue, Chicago, Illinois 60637, USA}}
\newcommand{\Adler}{\affiliation{Adler Planetarium, 1300 South DuSable Lake Shore Drive, Chicago, IL, 60605, USA}}
\newcommand{\CIERA}{\affiliation{Center for Interdisciplinary Exploration and Research in Astrophysics (CIERA), Northwestern University, 1800 Sherman Avenue, Evanston, IL, 60201, USA}}
\newcommand{\UNC}{\affiliation{Department of Physics and Astronomy, University of North Carolina at Chapel Hill, 120 E. Cameron Ave, Chapel Hill, NC, 27514, USA}}
\newcommand{\KICP}{\affiliation{Kavli Institute for Cosmological Physics, The University of Chicago, 5640 South Ellis Avenue, Chicago, Illinois 60637, USA}}
\newcommand{\EFI}{\affiliation{Enrico Fermi Institute, The University of Chicago, 933 East 56th Street, Chicago, Illinois 60637, USA}}
\newcommand{\UChicagoAA}{\affiliation{Department of Astronomy and Astrophysics, The University of Chicago, 5640 South Ellis Avenue, Chicago, Illinois 60637, USA}}
\newcommand{\CMU}{\affiliation{McWilliams Center for Cosmology and Astrophysics, Department of Physics, Carnegie Mellon University, Pittsburgh, PA 15213, USA}}

\newcommand{\COS}{\texttt{COSMIC}}
\newcommand{\MET}{\texttt{COSMIC-METISSE}}
\newcommand{\POS}{\texttt{POSYDON}}

\begin{document}

\title{When the stars don’t align: \\Investigating inconsistencies in binary black hole formation across population synthesis codes}

\author[orcid=0000-0002-8304-0109]{Alexandra G. Guerrero}
\UChicago
\email[show]{aghanselman@uchicago.edu}  

\author[orcid=0000-0002-0147-0835]{Michael Zevin}
\Adler
\CIERA
\email{mzevin@adlerplanetarium.org}

\author[0009-0000-8189-9887]{Duncan B. Maclean}
\UNC
\email{dmaclean@unc.edu}  

\author[0000-0001-5228-6598]{Katelyn Breivik}
\CMU
\email{kbreivik@andrew.cmu.edu}   

\author[0000-0003-4175-8881]{Carl L. Rodriguez}
\UNC
\email{carl.rodriguez@unc.edu} 

\author[0000-0002-6842-3021]{Max M. Briel}
\affiliation{Département d’Astronomie, Université de Genève, Chemin Pegasi 51, CH-1290 Versoix, Switzerland}
\affiliation{Gravitational Wave Science Center (GWSC), Université de Genève, CH1211 Geneva, Switzerland}
\email{max.briel@gmail.com}

\author[0000-0002-0175-5064]{Daniel E. Holz}
\UChicago
\EFI
\UChicagoAA
\KICP
\email{holz@uchicago.edu}

\begin{abstract}
Binary population synthesis (BPS) codes are valuable tools for investigating the end-to-end lives of binary stars as well as a myriad of observed astrophysical phenomena. Many ``rapid" BPS codes rely on semi-analytical single-star evolutionary tracks and disjointed prescriptions for binary physics. Updated ``hybrid" or ``detailed" BPS codes incorporate improved methodologies, but at a steeper computational cost that may limit broad exploration of physical uncertainties. Given the widespread use of BPS in modern astrophysical research, it is imperative to systematically compare BPS codes across the spectrum of computational efficiency, flexibility, and physical realism to gauge their consistency and robustness. In this work, we perform BPS on three single-metallicity populations of identical initial binaries using three modern codes --- the rapid code \texttt{COSMIC}, the hybrid code \texttt{METISSE} integrated into \texttt{COSMIC}, and the detailed code \texttt{POSYDON} --- ensuring consistent physical parameterizations where possible. We find stark differences in the properties, formation pathways, and progenitors of binary black hole (BBH) merger populations across the three codes. In a population of one million binaries at $0.01 Z_\odot$, each code results in $\sim 5,000\mbox{--}9,000$ BBHs that merge within a Hubble time. However, $\lesssim 14\%$ of progenitors overlap between two codes, and only \textit{one} initial binary becomes a BBH merger across all three codes. Identical progenitors often follow different evolutionary pathways in each code and result in inconsistent BBH properties. Our results highlight the need for a deeper understanding of physical and computational differences between BPS codes and caution against over-interpreting results from any individual BPS code.
\end{abstract}

\section{Introduction}
Binary systems containing massive stars are in the midst of an observational renaissance. Modern surveys are accumulating populations of systems and transient events that probe various stages in the lives of massive stars, such as supernovae with the Vera C. Rubin Observatory's Legacy Survey of Space and Time (LSST; ~\citealt{LSST_gen}), gamma-ray bursts with the Neil Gehrels Swift Observatory~\citep{2004ApJ...611.1005G,2016ApJ...829....7L}, detached binaries containing a compact object with Gaia~\citep{Gaia_DR3,2024A&A...686L...2G,2023MNRAS.521.4323E,2023MNRAS.518.1057E,2024NewAR..9801694E,2024OJAp....7E..58E,2026ApJ..1000..148L}, and compact binary mergers with the LIGO, Virgo, and KAGRA (LVK) \ac{GW} observatories~\citep{GWTC5_catalog,GWTC5_pop}.

The details of how binaries, particularly those containing massive stars, evolve are still highly uncertain. For example, processes such as stellar winds~\citep{2026arXiv260726147R}, the efficiency and stability of mass transfer~\citep{2024ARA&A..62...21M,2025ApJ...990L..51L,2026ApJ..1000....2S}, and the uncertainties in \ac{SN} physics and the properties of the compact remnants left behind~\citep{2021A&A...656A..58L,2024Univ...10..148B,2024ApJ...964L..16B,2026arXiv260104464F}, remain active areas of research. One way to constrain physical uncertainties in these systems is by computationally modeling large populations of binary stars through their whole evolution and comparing these simulations to observations. This approach, known as \ac{BPS}, has been a staple of binary-star studies over the past two decades and is an incredibly useful tool to study full, intrinsic populations over this large, uncertain physical parameter space~\citep{Breivik26}. Since \ac{BPS} codes allow for a direct mapping between physical quantities---for example, the \ac{ZAMS} to remnant mass function---they have been used as a blueprint to learn about a variety of astrophysical populations, such as detached star-compact object systems~\citep{2020ApJ...905..134W,2022MNRAS.511.5462T,2024A&A...692A.141K,2025PASP..137d4202N,2025A&A...704A...6R}, electromagnetic transient populations such as gamma-ray bursts~\citep{2006ApJ...648.1110B,2020MNRAS.491.3479C,2022A&A...657L...8B,2026A&A...710A.388D}, and \acp{GW} sourced from the merger of compact objects~\citep{2021ApJ...910..152Z,2022ApJ...933...86Z,2024MNRAS.534.3506R,2025ApJ...981...66D,2026arXiv260628515B,Maclean26,Chen2026,Briel26,Breivik26}.

Numerous \ac{BPS} codes have been developed over the past few decades. These \ac{BPS} codes each take different approaches in simulating the stellar and binary physics necessary to evolve binary stellar populations, but can be sorted into three broad categories. Rapid \ac{BPS} codes often use semi-analytic prescriptions or fitting formulae for stellar models to evolve each star in the binary, accounting for binary physics separately. Examples of rapid \ac{BPS} codes include \texttt{BSE}~\citep{Hurley_BSE}, \texttt{binary$\_$c}~\citep{2004MNRAS.350..407I,2006A&A...460..565I,2009A&A...508.1359I,2018MNRAS.473.2984I,2023MNRAS.521...35I,2023JOSS....8.4642H}, \texttt{COMPAS}~\citep{2017MNRAS.471.2801S,2018MNRAS.477.4685B,2022ApJS..258...34R,2025ApJS..280...43T}, \texttt{MOBSE}~\citep{2018MNRAS.480.2011G,2018MNRAS.474.2959G}, \texttt{SeBa}~\citep{1996A&A...309..179P,2001A&A...368..939N,2012A&A...546A..70T}, \texttt{StarTrack}~\citep{2002ApJ...572..407B,2008ApJS..174..223B,2016Natur.534..512B}, and \texttt{COSMIC}~\citep{COSMIC}. These rapid codes all use fitting formulae to the Single-Star Evolution (SSE) models given by \citet{Pols1998} for stellar evolution. Another set of \ac{BPS} codes integrate more modern stellar evolution tracks. Stellar evolution codes such as \texttt{MESA}~\citep{Paxton2011, Paxton2013, Paxton2015, Paxton2018, Paxton2019, Jermyn2023} or \texttt{BEC}~\citep{2000ApJ...528..368H,2000ApJ...544.1016H} fully evolve the differential equations governing single-star evolution. BPS codes such as \texttt{ComBinE}~\citep{2018MNRAS.481.1908K}, \texttt{SEVN}~\citep{2017MNRAS.470.4739S,2019MNRAS.485..889S,2023MNRAS.524..426I}, and \texttt{METISSE}~\citep{METISSE_comp, METISSE_bin, METISSE_JOSS} use pre-computed grids of single stars evolved with these stellar codes, and interpolate these grids to compute the stellar evolution of each star in the binary, with binary physics still added separately. The last category of \ac{BPS} codes, such as \texttt{BPASS}~\citep{2017PASA...34...58E,2018MNRAS.479...75S} and \texttt{POSYDON}~\citep{POSv1,POSv2}, use pre-computed grids of binary stars, self-consistently evolving the stellar and binary physics of each binary. These three categories of codes correspond to the ``rapid," ``hybrid," and ``detailed" \ac{BPS} codes, respectively, in ~\citet{Breivik26}, in which a more detailed summary of the differences between these codes can be found.

Rapid, hybrid, and detailed \ac{BPS} codes each provide different benefits. Although rapid BPS codes approximate much of the stellar and binary physics, they allow for an efficient and broad exploration of various physics assumptions. Hybrid \ac{BPS} codes treat stellar physics more robustly, but the approximate binary physics still allows for a quick exploration of physics assumptions. Finally, detailed \ac{BPS} codes provide a more precise view of stellar and binary physics, but only allow for limited explorations of the parameter space of physical assumptions. Even with the advancements in our understanding of binary star evolution, several studies have determined that \ac{BPS} codes are far from robust, often yielding different results across codes and prescriptions~\citep{2014A&A...562A..14T,METISSE_comp,2022ApJ...925...69B,2022ApJ...937..118W,2023MNRAS.525..706R,2023MNRAS.524..426I,2026arXiv260605322B,2026arXiv260628515B,2026arXiv260727391V}. Recent advancements in detailed binary grids have shown drastic differences between rapid and detailed codes, specifically around mass transfer stability~\citep{2017MNRAS.465.2092P,2021ApJ...922..110G,2023NatAs...7.1090B,2023MNRAS.520.5724B,2026A&A...706A.296K}. However, since \ac{BPS} codes are typically not well-calibrated, it is difficult to establish exactly where these discrepancies originate from. Nevertheless, building an understanding of uncertainties in \ac{BPS} output requires not just variations within a single \ac{BPS} codebase, but also thorough and systematic comparisons between codebases. Understanding where differences between \ac{BPS} codes come from is crucial for determining where improvements are required.

\citet{2014A&A...562A..14T} previously compared outputs between four rapid BPS codes, focusing specifically on white dwarf populations. They found good agreement in the progenitors, evolutionary pathways, and characteristics of these white dwarf populations, and found that the differences between codes were due to physical assumptions. In this work, we instead perform a detailed code comparison between the three broad categories of \ac{BPS} codes: the ``rapid" \ac{BPS} code \COS{}, the ``hybrid" single-star interpolation code \texttt{METISSE} integrated into \COS{} for binary evolution, and the ``detailed" binary star interpolation code \POS{}. By simulating binary stars with identical properties at birth and ensuring consistent physical choices wherever possible, we systematically investigate how differences in the treatment of stellar and binary physics between these codes affect the resultant binary evolution and compact object formation. Finally, taking the population of \ac{BBH} mergers as a case study, we demonstrate how these discrepancies manifest in final astrophysical populations, elucidating the uncertainties inherent in \ac{BPS} codes due to simulation techniques and physical prescriptions. 

The rest of the paper is organized as follows: In Section~\ref{sec:codes}, we describe the \ac{BPS} codes we use in this analysis. In Section~\ref{sec:methodDetails}, we define our method for implementing a consistent comparison between the codes. We then investigate the discrepancies in stellar properties for single stars and stars in binaries in Section~\ref{sec:results}. Finally, we examine discrepancies specifically in \ac{BBH} merger populations in Section~\ref{sec:pops}. We comment on the effects these differences have on \ac{GW} populations and discuss caveats to our work in Section~\ref{sec:discussion}, and conclude in Section~\ref{sec:conclusions}.

\section{Comparing binary population-synthesis codes\label{sec:methods}}
In this work, we synthesize identical \ac{ZAMS} populations of binary stars that are then evolved using each of three \ac{BPS} codes: \COS, \texttt{METISSE} integrated into \COS{} (hereafter denoted simply as \MET{}, and \texttt{METISSE} in figures as a shorthand), and \POS. We choose the evolutionary prescriptions such as to make each code evolve the same binaries in as similar a manner as possible. First, we describe the details of each code in Section~\ref{sec:codes}, before outlining how we generate the initial binaries and choose the evolutionary prescriptions in Section~\ref{sec:methodDetails}. 

\subsection{Population synthesis codes used in this work\label{sec:codes}}
Here, we provide a brief overview of each \ac{BPS} code, and point readers to references that provide a more complete description of each code. 
\subsubsection{\COS \label{sec:COS}}
\COS{} is a rapid \ac{BPS} code that evolves single stars following \citet{Hurley_SSE}, which implements fitting formulae to the SSE models given by \citet{Pols1998}. Binary interactions are based on BSE~\citep{Hurley_BSE}. \COS{} contains a number of modifications in both the treatment of single-star evolution and binary physics, such as updates to stellar winds, mass transfer stability, \ac{CE} evolution, remnant masses, and supernova kicks. See~\cite{COSMIC} for more information about the specific updates to BSE. In this work, we use \texttt{COSMIC v4.1.0}.\footnote{\texttt{COSMIC v4.1.0} is available on Zenodo~\citep{COSMIC_20721229}.}

\subsubsection{\MET \label{sec:MET}}
\texttt{METISSE}, or MEthod of Interpolation for Single Star Evolution, takes in a grid of pre-computed single star tracks evolved with any stellar-evolution code, as long as the tracks are formatted in Equivalent Evolutionary Point (\texttt{eep}) format~\citep{METISSE_bin}. In this work, we use the stellar tracks from the stellar-evolution code \texttt{MESA} provided in~\citet{Maclean26}.\footnote{We use the ``standard" tracks described in~\citet{Maclean26}. The \texttt{eep} files are publicly available at~\citet{Maclean_eeps}.} Given a star's initial mass, \texttt{METISSE} interpolates between the pre-computed tracks to evolve the star and re-interpolates the stellar characteristics based on any mass loss during \ac{RLOF}. Notably, during the post-main sequence phase, \texttt{METISSE} assumes that core evolution is not affected by mass loss in the envelope \citep{METISSE_bin}. \texttt{METISSE} can be integrated into any rapid \ac{BPS} code. We choose to integrate \texttt{METISSE} into \COS{} (hence \MET{}) for this work to enforce the same binary physics between \COS{} and \MET{}. Therefore, \MET{} replaces SSE for the stellar evolution of each star in a binary, but continues to add binary interactions following BSE, with the same modifications described in Section~\ref{sec:COS} and~\citet{COSMIC}. For further details, see \citet{METISSE_bin,METISSE_JOSS}. We additionally use updated stellar wind prescriptions to match those implemented in \POS{}~\citep{POSv2,Maclean26}. In this work, we use \texttt{METISSE} \texttt{v1.0.1}\footnote{\texttt{METISSE v1.0.1} is available on Zenodo~\citep{METISSE_17650929}.} as implemented in \COS{} \texttt{v4.1.0}.

\subsubsection{\POS}
\POS{} uses pre-computed grids of binary stars generated with \texttt{MESA}, where the internal structure of each star is evolved concurrently and self-consistently with the binary evolution. These pre-computed grids include \texttt{MESA} simulations of detached binaries and grids where the binaries interact through stable mass transfer, with each grid determining the evolution of a specific phase of the binary. Given a population of initial binary stars, \POS{} interpolates the binary stellar tracks from the pre-computed grids to follow the evolution of the binaries. The binary grids follow the evolution until core-carbon depletion of either star (generally the primary) or until an instability in the mass transfer is reached \citep[see][for the instability criteria]{POSv1, POSv2}. \ac{CE} evolution and \ac{SN} prescriptions are then calculated on-the-fly using detailed information from the stellar profile before continuing through the \texttt{MESA} grids. For a more detailed description of \POS{}, see \citet{POSv1,POSv2}. In this work, we use \texttt{POSYDON v2.2.2}.\footnote{\texttt{POSYDON v2.2.2} is available at: \url{https://github.com/POSYDON-code/POSYDON/releases}. The pre-computed grids are available at: \url{https://zenodo.org/records/15194708}.}

\subsection{Methods for consistent comparison\label{sec:methodDetails}}
As outlined in Section~\ref{sec:codes}, the methods used to evolve stellar binaries are different between each code. Even so, there are a number of prescriptions that can be made exactly equivalent between codes. When this is not possible, we attempt to choose prescriptions that are as similar as possible. We first begin by describing the method for generating our initial binary populations in Section~\ref{sec:initbins} before further describing the evolutionary prescriptions used for each code in Section~\ref{sec:prescriptions}. We also provide an overview of the prescriptions chosen in Appendix~\ref{sec:pres-table}. We stress that this paper focuses specifically on comparing these \ac{BPS} codes using as physically-consistent prescriptions as possible between codes. Therefore, we do not attempt to consider all possible variations in binary physics. However, we do explore two settings available in the ``rapid" BPS framework \COS{} in Appendix~\ref{sec:bin-variations} to demonstrate the large impact unconstrained physics can have on final \ac{BBH} populations. 

\subsubsection{\label{sec:initbins}Generating initial binaries}
 An initial population of binary stars is generated using \POS{} for three fiducial metallicities, $\{0.01, 0.1, 1\} Z_\odot$, with one million binaries in each single-metallicity population. In this work, we use $Z_\odot = 0.0142$ to be consistent with~\citealt{2009ARA&A..47..481A,POSv2}. The initial masses of each primary star, $m_{1, \rm ZAMS}$, are drawn from a Kroupa initial mass function~\citep[IMF;][]{2001MNRAS.322..231K} between 7 and 150 $M_\odot$. The masses of the companions are drawn using a flat-in-mass-ratio distribution with secondary masses, $m_{2, \rm ZAMS}$, spanning from 0.35 to 150 $M_\odot$. We require that the primary star, denoted everywhere by a 1 subscript, is always initially more massive than its companion, i.e. $m_{1, \rm ZAMS} \geq m_{2, \rm ZAMS}$. Initial orbital periods, $p_{\rm orb, ZAMS}$, are drawn following the extension to~\citet{2012Sci...337..444S} found in Appendix A of~\citet{2021A&A...647A.153B}: for primary masses greater than $15 M_\odot$, the orbital period distribution is a power law with index $-0.55$ for periods greater than $10^{0.15}$ and a flat-in-log distribution for periods less than $10^{0.15}$; for primary masses less than $15 M_\odot$, all orbital periods are drawn from a flat-in-log distribution. We draw orbital periods in the range of 0.75 to 6,000 days. All binaries are generated with zero initial eccentricity for simplicity of comparison, although eccentricity is allowed during evolution. After these initial binary populations are generated, they are fed into \COS{} and \MET{} as inputs to ensure that we are able to track each individual binary as it is evolved with each code. Note that we generate all binaries at the same epoch, and evolve the systems for a Hubble time, which we set to $13.7$~Gyr, to be consistent across codes. To standardize the orbital decay due to \ac{GW} emission, we select all binaries that form \acp{BBH} and calculate the \ac{GW} inspiral time $t_{\rm GW}$ using~\citet{Peters1964}. We then define binaries that become \ac{BBH} mergers as those with a delay time $t_d \equiv t_{\rm DCO} + t_{\rm GW} < 13.7$~Gyr, where $t_{\rm DCO}$ is the time from birth to \ac{DCO} formation. In this work, we focus on the properties of these merging \ac{BBH} systems at \ac{BBH} formation. We additionally turn off \ac{SN} kicks to ensure their inherent randomization do not have an impact on whether or not a particular system reaches a particular end state in one code versus another. 

\subsubsection{\label{sec:prescriptions}Choice of evolutionary prescriptions}
Below, we provide brief descriptions of the specifics of the evolutionary prescriptions; we point readers to \citet{COSMIC}, \citet{METISSE_bin} and~\citet{Maclean26}, and \cite{POSv2} for specific details on how these prescriptions are implemented in \COS, \MET, and \POS, respectively. Naturally, there are some evolutionary prescriptions that are unable to be exactly replicated between the codes. We highlight the key differences in these prescriptions below, although we strive to make these differences as small as possible. Table~\ref{tab:prescriptions} in Appendix~\ref{sec:pres-table} also provides a schematic overview of the prescription choices used in each code.

\textit{Stellar wind prescriptions:}
\COS{} stellar winds are taken from~\citet{Vink2001} for O/B stars, with B stars having an effective temperature $12,500 \rm K < T_{\rm eff} < 22,500 \rm K$ and O stars having $27,500 \rm K < T_{\rm eff} < 50,000 \rm K$. The prescription in~\citet{2013ApJ...779...72D} is used to extend winds for both O and B stars to an effective temperature of $25,000 \rm K$. Wolf-Rayet (WR) stars are treated using~\citet{Vink2005}. Luminous blue variable (LBV) winds from~\citet{2010ApJ...714.1217B} are used with a constant mass loss rate of $1.5 \times 10^{-4} M_\odot \text{yr}^{-1}$ for stars beyond the Humphreys-Davidson (HD) limit~\citep{1979ApJ...232..409H}. See~\citet{COSMIC} for a more detailed description of the specific wind prescriptions. 

\MET{} and \POS{} have the same stellar wind prescriptions. For stars with $M_{\rm ZAMS} > 8 M_\odot$, winds are treated using the \texttt{MESA Dutch} scheme, which uses (``cool") winds from~\citet{deJager1988} for stars with $T_{\rm eff} < 10,000 \rm K$. For stars with $T_{\rm eff} > 11,000 \rm K$, (``hot") winds from~\citet{Vink2001} are used if the surface hydrogen mass fraction, $X_{\rm surf}$, is greater than 0.4. Otherwise, the winds for these stars are replaced with WR winds from~\citet{Nugis2000}. Stars with $M_{\rm ZAMS} > 8 M_\odot$ and effective temperatures between $10,000 \rm{K} < T_{\rm eff} < 11,000 \rm{K}$ have winds that are a linear interpolation of the two prescriptions. For stars in \MET{} and \POS{} with $M_{\rm ZAMS} < 8 M_\odot$ and $T_{\rm eff} > 12,000 \rm K$, the \texttt{MESA Dutch} scheme is also used. For stars with $T_{\rm eff} < 8,000 \rm K$, the prescription of~\citet{1975psae.book..229R} is used with a scaling factor $\eta_R=0.1$ for the first ascent of the giant branch (GB). Otherwise, the prescription of~\citet{Bloecker1995} is used with $\eta_B=0.2$ for stars in the thermally pulsating phase. Stars with $8,000 \rm K < T_{\rm eff} < 12,000 \rm K$ have winds that are linearly interpolated between the hot and cool schemes. LBV winds are calculated using the prescription from~\citet{2010ApJ...714.1217B} for stars that cross the HD limit, with a mass loss rate of $10^{-4} M_\odot \rm yr^{-1}$.  
See Figure 3 in~\citet{POSv2} for a schematic of how winds are implemented in \POS{} and Section 2.1.1 in~\citet{Maclean26} for a description of how these winds are implemented in \MET{}.

While stellar winds in \COS{} are calculated as a post-processing step to stellar evolutionary models that originally were wind free, winds in \MET{} and \POS{} are calculated during the star's evolution.

\textit{Mass transfer and \ac{CE} prescriptions:} Mass transfer occurs between a donor and accretor when the donor's radius overfills its Roche lobe and initiates \ac{RLOF}. In \COS{} and \MET{}, the mass loss rate of the donor is calculated following Equations 58 and 60  in~\citet{Hurley_BSE} for nuclear timescales and thermal timescales, respectively. The amount of this material that the accretor accumulates is limited to ten times the thermal rate of the accretor for main-sequence (MS), Hertzsprung gap (HG), and core He burning (CHeB) stars, but is unlimited for giant-branch (GB), early asymptotic giant-branch (EAGB), and asymptotic giant-branch (AGB) stars. For degenerate companions, the mass transfer is limited by the Eddington rate. \POS{}, on the other hand, uses the \texttt{contact} scheme in \texttt{MESA} for MS stars, and the \texttt{Kolb} scheme~\citep{Kolb1990} for post-MS stars with a central H abundance less than $10^{-6}$~\citep{POSv1}. At the beginning of \ac{RLOF} with a non-degenerate companion, all mass lost from the donor is accreted and transfers angular momentum following~\citet{2013ApJ...764..166D}, leading to the rapid spin-up of the accretor. If the accretor becomes critically rotating, the accretion becomes inefficient (see, however, ~\citet{2022A&A...659A..98S,2024ApJ...971..133R,2025A&A...704A.218X,Briel26} for examples of efficient mass transfer in short-period binary systems due to tidal spin-down). If the accretor is a degenerate star, the mass transfer is capped at the Eddington-limited rate~\citep{POSv1}. In all three codes, any excess mass lost from the donor that is not accreted by its companion is lost as if it is a wind from the accretor, and carries the specific angular momentum of the accretor. 

In all three codes, binaries are assumed to go through a \ac{CE} phase if mass transfer becomes unstable. However, the details of when this transition occurs varies between the codes. For \COS{} and \MET{}, mass transfer becomes unstable if the mass ratio, $m_{\rm donor}/m_{\rm accretor}$, exceeds a critical mass ratio, $q_{\rm crit}$. The values for $q_{\rm crit}$ depend on the phase of the donor, but are detailed in \cite{1987ApJ...318..794H} for GB/AGB stars, and in \cite{Hurley_BSE} for all other donors (\texttt{qcflag=1}). However, there are also cases in \COS{} and \MET{} where mass transfer is always unstable regardless of mass ratio. This happens when either the stellar radii of both the donor and accretor go into contact with each other or if the donor's radius overflows its component's Roche radius~\citep{COSMIC}. However, we set the \COS{} flag \texttt{smt$\_$periastron$\_$check} to off such that \COS{} and \MET{} do not explicitly check if the periapse distance of the orbit is less than either stars' radii\footnote{This check was implicitly enforced starting in \texttt{COSMICv3.6}, but became a flag with the option to turn it off starting in \texttt{COSMICv3.7.8}.} to better match with \POS{}. In \POS{}, mass transfer is deemed unstable and a \ac{CE} phase is triggered if either the mass transfer rate exceeds $0.1 M_\odot \rm yr^{-1}$, if the donor's radius overflows the second Lagrange point (L$_2$), if both stars overflow their Roche lobes while one of the stars is off the main sequence (MS), or if the photon trapping radius exceeds the Roche lobe of the accretor when the accretor is a compact object~\citep{POSv1}. Therefore, while \COS{} and \MET{} start a \ac{CE} event instantaneously if the above conditions are met, \POS{} has a period of stable mass transfer before a \ac{CE} event starts.

Once a \ac{CE} phase starts, all three codes use the $\alpha$--$\lambda$ prescription following~\citet{Hurley_BSE}. The \ac{CE} efficiency is set to $\alpha=1.0$ for all three codes. A \ac{CE} is successful if there is enough orbital energy in the system to overcome the envelope binding energy and eject the envelope before the inspiraling objects merge. In \COS{}, the envelope binding energy factor is computed as in~\citet{Hurley_BSE} and~\citet{Claeys2014}. However, as \MET{} and \POS{} have precise knowledge of the density profile of the star, the binding energy is computed directly by integrating the gravitational energy of the envelope~\citep{Maclean26,POSv1}. In both \MET{} and \POS{}, we set the core during the \ac{CE} phase to be defined  as the outermost layer where the H fraction drops below $10\%$. Note, however, that \POS{} typically has already undergone some mass transfer before the \ac{CE} starts, and thus will have some structural changes compared to \MET{} systems. If the envelope ejection criteria is not met, the donor and accretor merge in the CE. If the criteria is met, the envelope is successfully ejected and the binary continues to evolve. In \POS{}, we use the \texttt{one$\_$phase$\_$variable$\_$core$\_$definition} flag such that there is no mass transfer after a successful \ac{CE} ejection to be consistent with \COS{} and \MET{}. In this work, we use the optimistic \ac{CE} prescription for all codes, where HG donors are allowed to survive a \ac{CE} phase~\citep{2008ApJS..174..223B}. 

\textit{\ac{SN} and remnant mass prescriptions:} When a star reaches the end of its life, \ac{SN} prescriptions are used to calculate the mass of the remnant compact object. In \COS{} and \MET{}, stars are evolved until core-He depletion, whereas \POS{} evolves stars until core-carbon depletion. All three codes then use the delayed prescription from~\citet{2012ApJ...749...91F} to calculate the remnant mass of the compact object. Stars with sufficiently massive cores can instead go through pulsational pair instabilities (PPIs), where spontaneous burning in the star's core due to the creation of electron-positron pairs causes material to be ejected from the star's surface~\citep{2007Natur.450..390W,2015ASSL..412..199W,2017ApJ...836..244W}. Stars with core masses above this mass range can go through \ac{PISN}, where no compact object is left behind. All three codes use the (P)\ac{PISN} prescription in~\citet{Hendriks:2023yrw} based on the top-down approach in~\citet{2022RNAAS...6...25R}, with no extra mass loss or shifts to the C/O core mass PPI threshold added. These codes also use an \ac{ECSN} prescription for low-mass stars that never develop an iron core~\citep{2004ApJ...612.1044P}. Although in this work we focus on black hole formation, which is typically outside the mass range of stars that go through \ac{ECSN}, we report the prescriptions used in the three codes here for completeness. \POS{} determines if a star goes through an \ac{ECSN} if its core mass is within the range provided by~\citet{2015MNRAS.451.2123T}. In \COS{} and \MET{}, we choose stars to undergo an \ac{ECSN} if their He-core mass is between $(2.6-2.95) M_\odot$ at the end of core helium burning, as an approximation for the prescription outlined in~\citet{2015MNRAS.451.2123T}. 

For the main results in this work, we do not include \ac{SN} kicks to reduce any aspect of randomness. However, additional results including \ac{SN} kicks are shown in Appendix~\ref{sec:kicks}. The details of the kick prescriptions used for those results are described therein.

\section{Evaluating the impact of stellar and binary physics across BPS codes\label{sec:results}}
\begin{figure*}[tbh]
    \includegraphics[width=\textwidth]{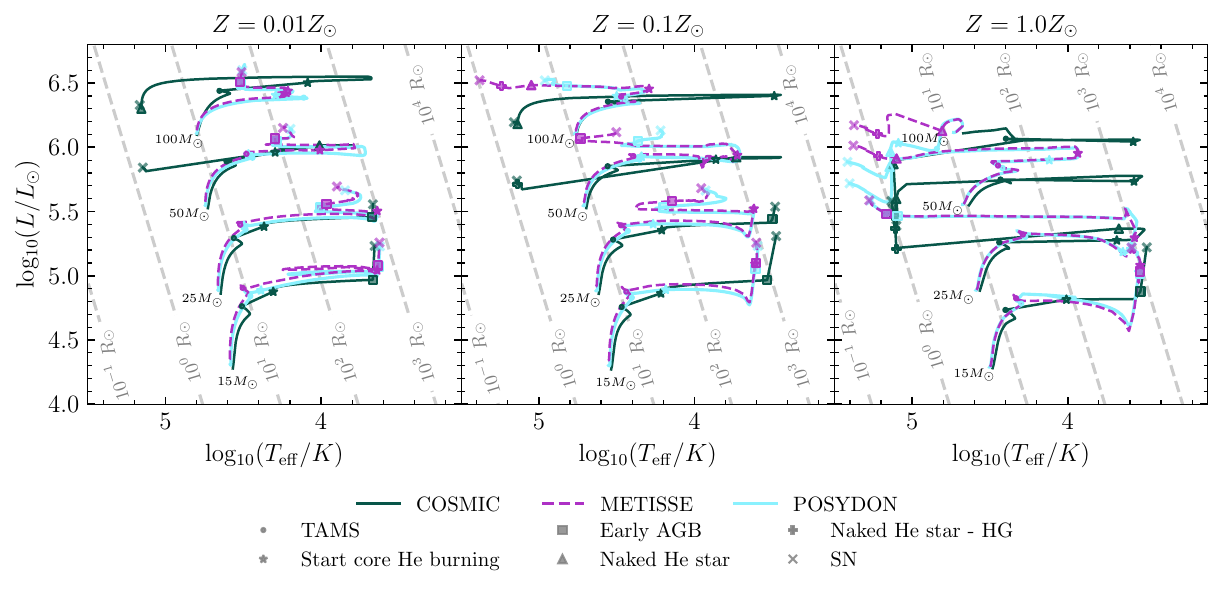}
    \caption{HR diagrams for single stars evolved using \COS{} (dark green), \MET{} (magenta dashed), or \POS{} (light blue) for three metallicities: $0.01 Z_\odot$ (left panel), $0.1 Z_\odot$ (middle panel), and $1.0 Z_\odot$ (right panel). The exact \ac{ZAMS} masses for the stars in the above tracks are $\{14.98$, $ 25.04$, $ 49.68$, $ 100.69\} M_\odot$ to match exact grid points in \POS's \texttt{MESA} tracks. Markers indicate important steps in the star's evolution. Gray dashed lines denote stars with a constant radius.}
    \label{fig:HRdiagram}
\end{figure*}

Differences in prescriptions and simulation techniques between \ac{BPS} codes will have a substantial effect on the lives and deaths of massive stars. Here, we analyze the stellar properties of single and binary stars evolved with \COS{}, \MET{}, and \POS{}. We first investigate deviations in the lifecycles of massive single stars in Section~\ref{sec:singleStars}. We then isolate discrepancies in radial expansion in Section~\ref{sec:rmax} and final compact object remnant masses in Section~\ref{sec:mrem}, as these parameters have large impacts on final \ac{BBH} populations. 

\subsection{Comparison of single star evolution\label{sec:singleStars}}

To  isolate which differences are due to discrepancies in stellar evolution as opposed to binary interactions, we first investigate the evolution of single stars. We evolve four stars, with \ac{ZAMS} masses of $\{14.98$, $ 25.04$, $ 49.68$, $ 100.69\} M_\odot$, for three fiducial metallicities, $\{0.01,0.1,1\} Z_\odot$. We choose these \ac{ZAMS} masses such that the stars lie on exact grid points in the \POS{} \texttt{MESA} grids. The evolutionary tracks of these single stars evolved with \COS{}, \MET{}, and \POS{} are shown in the \ac{HR} diagrams in Figure~\ref{fig:HRdiagram}. We also include important stages in the stars' evolutions as shaped markers. The stellar tracks for \MET{} and \POS{} are closely matched for all three metallicities. \COS, on the other hand, shows much larger discrepancies in the stellar tracks across all masses and metallicities, although there is less difference for smaller \ac{ZAMS} mass and lower metallicities. 

As discussed in Section~\ref{sec:prescriptions}, \texttt{MESA} is used for the stellar evolution in both \MET{} and \POS, and single-star prescriptions such as stellar winds are identical. The main mismatch between \MET{} and \POS{} comes from the fact that, although the \ac{ZAMS} masses for the stars in Figure~\ref{fig:HRdiagram} are chosen to lie on exact \texttt{MESA} \POS{} grid points, these do not lie on exact \texttt{MESA} grid points in \MET. Therefore, the \ac{HR} tracks in Figure~\ref{fig:HRdiagram} for \MET{} are an interpolation between the closest grid points in \MET, which can shift the tracks by a small amount. The later stages of the $50 M_\odot$ and $100 M_\odot$ tracks at $0.1 Z_\odot$ and $1 Z_\odot$ have more substantial deviations between \MET{} and \POS{}. These larger differences can be attributed to \POS{} using the \texttt{MESA} revision \texttt{r11701},\footnote{\texttt{MESA r11701} can be found on Zenodo~\citep{MESA_r11701}.} while \MET{} uses the \texttt{MESA} revision \texttt{r24.08.1}.\footnote{\texttt{MESA r24.08.1} can be found on Zenodo~\citep{MESA_13353788}.} Specifically, \texttt{r24.08.1} updated the \texttt{eos} tables~\citep{Jermyn2023,MESA_13353788}, and handles the convective core edge and extremely H-poor convective envelope differently. This leads to a slightly different core mass and stellar radius, which can impact the stellar luminosity, especially in these high-mass regimes~\citep{Paxton2019,Jermyn2023,2024ApJS..274...30Z}. However, we stress that the overall differences in stellar evolution between \MET{} and \POS{} are very minor for most masses and metallicities relevant in this work.

\COS{} uses a different stellar wind prescription than \MET{} and \POS{}. While all three codes follow~\citet{Vink2001} while the star is on the main-sequence, \MET{} and \POS{} transition to the cool wind scheme once the temperature falls below $10,000$~K. The cool winds from~\citet{deJager1988} are stronger than~\citet{Vink2001}, leading to a higher rate of mass loss in \MET{} and \POS{} such that tracks rapidly increase to hotter temperatures and decrease to smaller radii. Once the star loses enough of its hydrogen envelope, the winds transition to WR winds. \COS{} uses WR winds following~\citet{Vink2005} while \MET{} and \POS{} use~\citet{Nugis2000}. At $0.01 Z_\odot$, the WR winds in~\citet{Vink2005} are much stronger than those in~\citet{Nugis2000}, leading to more of the envelope being blown off in higher-mass systems in \COS{} and a sharp increase to hotter temperatures (see e.g. the markers showing the transitions to naked He stars). For the high-mass stars at all metallicites, while the \MET{} and \POS{} tracks contract to smaller radii due to the cool stellar winds, the same stars evolved with \COS{} expand to larger radii due to the weaker~\citet{Vink2001} winds, cross the HD limit, and transition to LBV winds. The transition to LBV winds in \COS{} causes the sharp increase to high temperatures and decrease to lower stellar radii at the later stages of these stars. Additionally, \COS{} also extrapolates stellar tracks above $50 M_\odot$, as the original tracks were only computed between 0.5 and $50 M_\odot$~\citep{Hurley_SSE}, which also contributes to the extreme disagreement between \COS{} and the other two codes in the $100 M_\odot$ tracks.

\subsection{Comparison of the maximum radial evolution of single and binary stars\label{sec:rmax}}
\begin{figure*}
    \includegraphics[width=\textwidth]{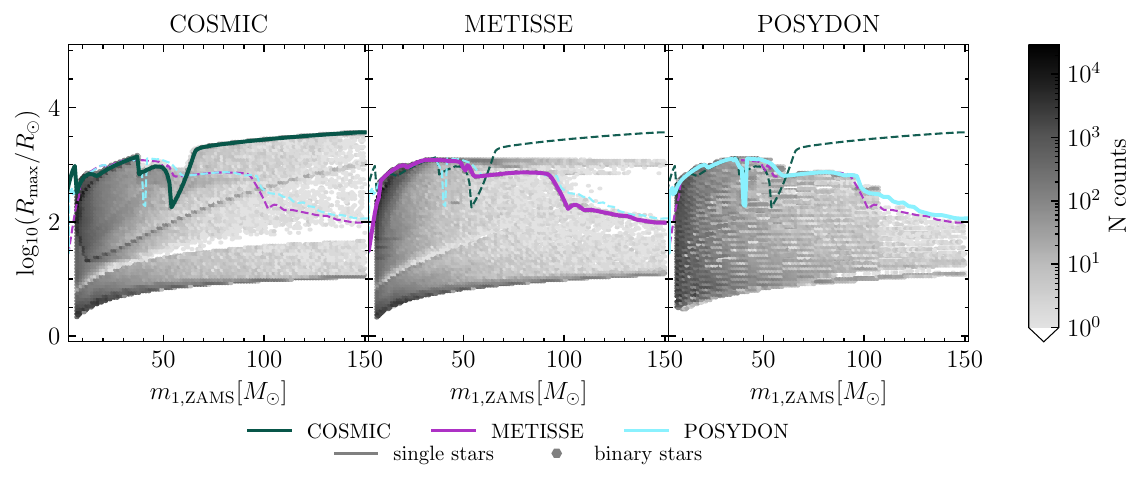}
    \caption{Maximum radii single stars (solid lines) expand to during their lives as a function of \ac{ZAMS} mass in \COS{} (left panel; dark green), \MET{} (middle panel; magenta), and \POS{} (right panel; light blue) for a fiducial metallicity of $Z=0.01 Z_\odot$. Dashed lines indicating the single star trends for the other two codes are provided as a reference in each panel. Taking our population of binary stars generated using the method described in Section~\ref{sec:initbins}, we bin all initially more massive stars according to their \ac{ZAMS} mass and maximum stellar radius (gray hexbins). The darker to lighter gradient from low to high \ac{ZAMS} mass is a result of the population's IMF. The large deviations from the single star line demonstrate the large effect binarity can have on a star's evolution.}
    \label{fig:Rmax}
\end{figure*}
The mismatch in stellar physics prescriptions in \COS{} as compared to \MET{} and \POS{} lead to large deviations in the evolution of massive stars. For example, \COS{} stars typically experience larger radial expansion during the post-main sequence phase, which will have a major impact on how these stars interact within a binary system. More specifically, the radial expansion will determine the moment and type(s) of mass transfer in binaries, thereby directly driving the characteristics of final astrophysical populations. 

Given the profound impact stellar radii can have on astrophysical populations, we now examine the systematic differences in stellar radii of both single stars and those in binaries between \ac{BPS} codes. In Figure~\ref{fig:Rmax}, we plot the maximum radius a single star expands to as a function of its \ac{ZAMS} mass as the solid lines, for a fiducial metallicity of $0.01 Z_\odot$. To investigate the effect binarity has on realistic binary populations, we additionally calculate the maximum radii for the initially more massive star in our $0.01 Z_\odot$ binary population and plot the distribution of maximum radii for this population as the gray hexbins.\footnote{For clarity, we bin the primary stars in our $10^6$ binary systems such that the colorbar indicates the number of binaries in our initial population that start at a given \ac{ZAMS} mass and report a specific maximum radius within the span of its life. As certain systems may merge before a given primary star fully evolves, we calculate the maximum radius of each primary star strictly before any stellar mergers occur. }

The maximum radii for single stars with \ac{ZAMS} mass $\lesssim 40 M_\odot$ match very well between all three codes. Furthermore, the maximum radii match up well between \MET{} and \POS{} for most \ac{ZAMS} masses. However, the two codes deviate slightly at higher \ac{ZAMS} mass due to the differences between \texttt{MESA} versions mentioned in Section~\ref{sec:singleStars}. The maximum radii in \COS{} deviate strongly for \ac{ZAMS} masses above $\sim 70 M_\odot$. The two notches in \COS{} stellar radii for \ac{ZAMS} masses between $40$ and $70 M_\odot$ are due to a sharp transition to WR and LBV winds. For \ac{ZAMS} masses above $\sim 70 M_\odot$, the maximum stellar radii in \COS{} are almost an order of magnitude larger than in \MET{} and \POS, and deviate to over an order of magnitude at larger \ac{ZAMS} masses. These deviations in maximum stellar radius for single stars between the three codes are directly related to the conclusions we found above in Section~\ref{sec:singleStars}, and are primarily driven by the difference in stellar winds and the extrapolation of the~\citet{Hurley_SSE} tracks in \COS{}.

Adding a stellar companion introduces binary interactions, with either stable or unstable mass transfer phases, which affect the amount the primary star can expand during its life. First, as our binary population follows a Kroupa IMF, there are de facto more binaries at smaller \ac{ZAMS} masses, illustrated by the gradient of dark to light bins as \ac{ZAMS} mass increases in Figure~\ref{fig:Rmax}. Next, as is shown by the spread in the maximum radii for a given \ac{ZAMS} mass, including binary interactions sometimes leads to drastically different maximum radii than those for single stars. This trend to lower maximum radii than a primary's single star counterpart is due to the primary interacting with its companion, bounding the amount the primary can expand before transferring mass to its companion. Note that the gradient in \POS{} binaries does not follow as smooth of a distribution as \COS{} and \MET{}. In \POS{}, fine timestep evolution of stellar parameters is currently only available through nearest neighbor interpolation. Therefore, in Figure~\ref{fig:Rmax}, we calculate the maximum radius of each star in the binaries by evolving the star's nearest neighbor equivalent in the \POS{} \texttt{MESA} grids, which leads to the coarse behavior seen in the maximum radii of \POS{} binary stars.

In all three codes, primary stars typically do not exceed the absolute maximum radii given by the single star trend, as expected. Note that there are some binary systems in \MET{} with maximum radii above the single star line for a given \ac{ZAMS} mass.
This is because in \MET{} (as well as \COS{}), stellar radii are not self-consistently recalculated during \ac{RLOF}; instead, the donor's radius is estimated as if it were a star of an equivalent mass that is not undergoing mass transfer. 
The primaries that have maximum radii above their single-star counterparts go through this ``track-jumping", losing enough mass during mass transfer to transition to a lower-mass \texttt{MESA} track that can reach a larger maximum radii.\footnote{Note that although \COS{} also experiences track jumping, the maximum radius for single stars is larger for high \ac{ZAMS} mass systems. Thus, when high-mass primaries lose mass in \COS{}, they still transition to a lower-mass stellar track, but this track does not reach a larger maximum radius.} This ``track-jumping" does not occur in \POS{} as the radius of each star in the binary is self-consistently calculated as mass is transferred. 

\begin{figure*}[tbh]
    \includegraphics[width=\textwidth]{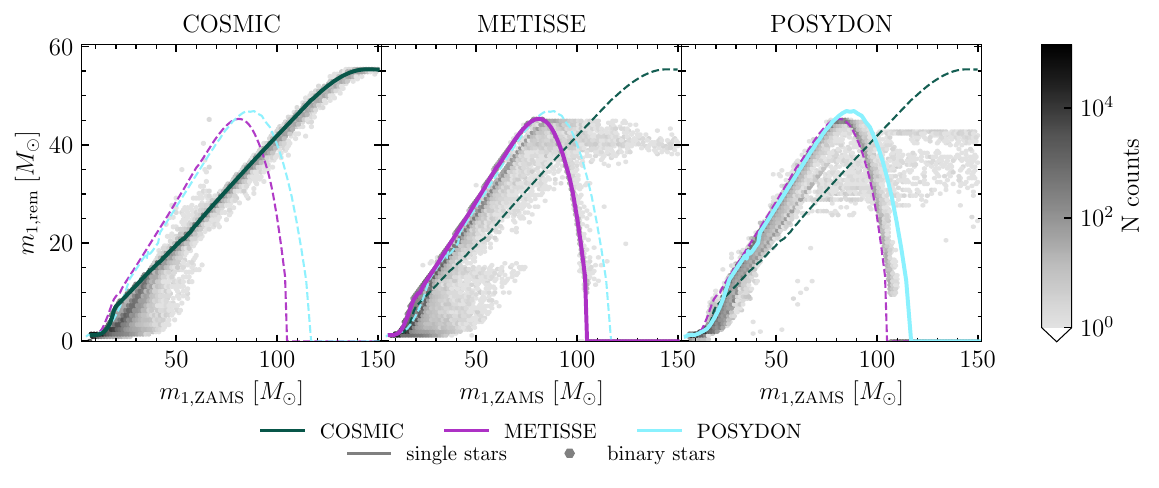}
    \caption{Remnant compact object mass of the primary star directly after \ac{SN} as a function \ac{ZAMS} mass for single stars (solid line) and for stars in binaries (gray hexbins) for a fiducial metallicity of $Z=0.01 Z_\odot$ in \COS{} (left panel; dark green), \MET{} (middle panel; magenta), and \POS{} (right panel; light blue). Dashed lines indicate the single-star trends for the other two codes as a reference. The binary stars are taken from our single-metallicity population evolved with prescriptions defined in Section~\ref{sec:initbins}. Note that the majority of binary systems typically follow the same trend as the single stars, with a broad offset dictated by the amount of mass transfer before \ac{SN}.}
    \label{fig:mrem}
\end{figure*}

Finally, note that the minimum radius is set by the radius a star has at \ac{ZAMS}. Primary stars that have their maximum radius at \ac{ZAMS} are in very short initial orbital periods such that the star cannot expand before interacting with its companion. Because our initial binary population has a \ac{ZAMS} orbital period distribution with a decaying power law, more systems have smaller orbital periods than larger ones at \ac{ZAMS}, hence the denser region at maximum radii close to the \ac{ZAMS} stellar radii.

\subsection{Comparison of compact object remnant properties\label{sec:mrem}}
Due to its effect on mass transfer phases, the maximum radial extent of the donor star ultimately determines the core mass available at the time of core collapse. When a star reaches the end of its life, \ac{BPS} codes use prescriptions to link the pre-supernova state of the star to compact object properties. The remnant mass of the compact object is determined by the final mass and C/O core mass of the star immediately before \ac{SN}~\citep[e.g.,][]{2012ApJ...749...91F}. Therefore, during mass transfer, both the amount of mass transferred and the distribution of mass within the star will affect the core mass and hence the remnant mass of the compact object.

To evaluate the cumulative impact of stellar and binary physics on final compact object properties, we plot the relation between the \ac{ZAMS} mass and remnant compact object of single (solid lines) and primary stars in binaries\footnote{The gray hexbins plot the remnant mass of the primary directly after core-collapse. Therefore, while Figure~\ref{fig:mrem} does track mass transfer phases before the first \ac{SN}, it does not track any mass the primary's compact object may accrete due to further stages of mass transfer. We also omit any primary stars that merge with its companion before core collapse.} (gray hexbins) in Figure~\ref{fig:mrem} for our fiducial $0.01 Z_\odot$ metallicity population. Similar to the results in Section~\ref{sec:rmax}, the single-star trends of \ac{ZAMS} to remnant mass are much more consistent between \MET{} and \POS{} than the single-star trend in \COS{}. As outlined in Section~\ref{sec:prescriptions} and Table~\ref{tab:prescriptions}, all three codes use the delayed remnant mass prescription from~\citealt{2012ApJ...749...91F} for core-collapse \ac{SN} and the (P)\ac{PISN} prescription from~\citealt{Hendriks:2023yrw}. However, while  \COS{} uses a prescription for the core-envelope boundary, \MET{} and \POS{} calculate the remnant mass from the C/O core mass, along with the amount of ejecta that falls back onto the compact object, directly from the stellar profile provided by \texttt{MESA}. The use of the~\citet{Hurley_SSE} tracks in \COS{} leads to significantly smaller C/O core masses. Therefore, \COS{} single stars do not hit the PPISN regime until much larger \ac{ZAMS} masses. On the other hand, the slight differences between \MET{} and \POS{} at $m_{1, \rm ZAMS} \gtrsim 80 M_\odot$ are mainly due to differences in stellar cores stemming from the different \texttt{MESA} revisions, as discussed in Section~\ref{sec:singleStars}. This difference leads to a slightly larger maximum black hole mass in \POS{}, around $47 M_\odot$, as opposed to $\sim 45 M_\odot$ in \MET{}, and also leads to a different maximum \ac{ZAMS} mass that collapses into a compact object, $\sim 105 M_\odot$ for \MET{} versus $\sim 118 M_\odot$ for \POS{}. 

Including a stellar companion again muddles the clear single-star trends seen in Figure~\ref{fig:mrem}. This additional scatter is directly due to the mass transfer episodes occurring in each binary. If we use the single star relationship as a reference and consider slices in primary \ac{ZAMS} mass, we see that the majority of primaries in all three codes almost always have remnant masses that are less than that of a single star of the same \ac{ZAMS} mass. This is due to mass transfer phases shrinking the mass of the envelope, leading to less fallback. There are, however, primaries in \MET{} and \POS{} with $m_{1, \rm ZAMS} \gtrsim 100 M_\odot$ that are above the lower (P)\ac{PISN} boundary that lose enough mass to avoid \ac{PISN}, forming a compact object when their single-star counterpart does not. Of the primaries with masses greater than their single-star counterparts in \COS{}, many undergo a reverse mass transfer phase before the first \ac{SN}. 

The disagreement in \ac{ZAMS}-to-remnant mass relations between the three codes further emphasizes that, even when given an identical initial binary population, any \acp{DCO} formed will have different masses depending on which code is used. Consequently, these deviations in stellar and binary physics prescriptions and treatments between \ac{BPS} codes will heavily influence the population-level properties of merging compact objects, such as \ac{BBH} mergers.

\section{Discrepancies in the populations of merging binary black holes\label{sec:pops}}
Having examined the variations in single and binary stellar properties across different \ac{BPS} codes, we now investigate the impact these differences have on the aggregate properties of the resulting \ac{BBH} merger populations. Here, we restrict our analysis to a fiducial metallicity of ${Z=0.01 Z_\odot}$. The corresponding $0.1 Z_\odot$ and $1.0 Z_\odot$ populations are provided in Appendix~\ref{sec:otherZ}, and the \ac{BBH} populations formed when including \ac{SN} kicks for a $0.01 Z_\odot$ fiducial metallicity is provided in Appendix~\ref{sec:kicks}.

\subsection{\ac{BBH} progenitor and merger populations diverge across codes\label{sec:scatters}}

\begin{figure*}
    \includegraphics[width=\textwidth]{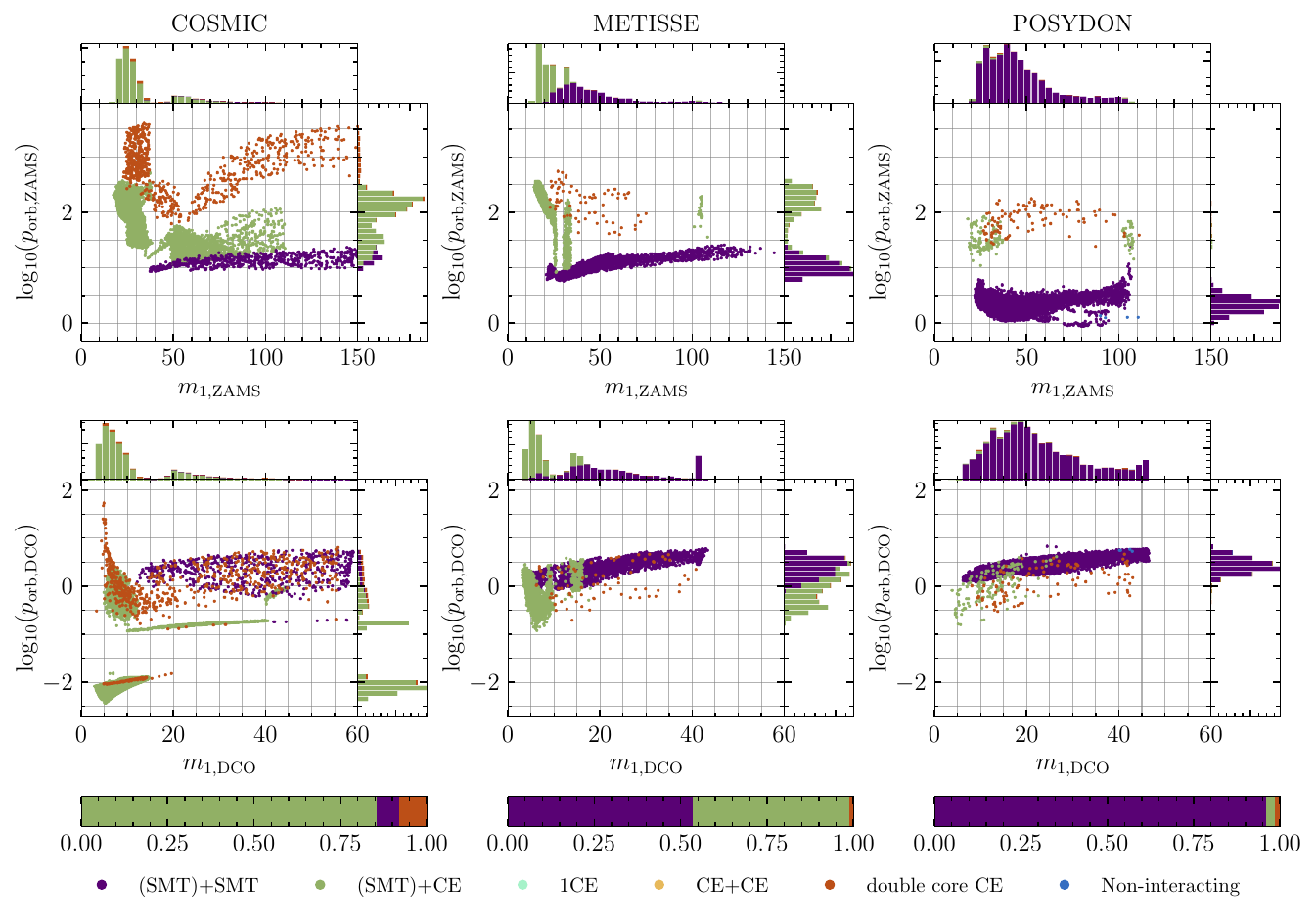}
    \caption{Primary masses and orbital periods at \ac{ZAMS} and \ac{BBH} formation, where the primary mass, $m_1$, is always defined as the mass of the initially more massive star at \ac{ZAMS}. (Top panels) Scatter points of individual \ac{ZAMS} binaries that end up becoming merging \ac{BBH}s when evolved with \COS{} (left panels), \MET{} (middle panels), and \POS{} (right panels), respectively, colored by mass transfer type, for a single metallicity of $Z = 0.01 Z_\odot$. The colors indicate if the system went through solely stable mass transfer either both before and after or only after the first \ac{SN} ((SMT)$+$SMT; purple), no or stable mass transfer before the first \ac{SN} with a \ac{CE} phase after the first \ac{SN} ((SMT)$+$CE; green), one \ac{CE} event before the first \ac{SN} (1CE; cyan), a \ac{CE} phase both before and after the first \ac{SN} (CE$+$CE; gold), if both stars begin unstable mass transfer at the same time (double core CE; dark orange), or if the binaries never go through a mass transfer event (non-interacting; navy blue). The respective one-dimensional marginal distributions are also shown on the top and right side in each panel as stacked histograms. (Bottom panels) Same as top panels, but now looking at properties of merging \acp{BBH} at the formation of the \ac{BBH}. Below these plots are the percentage of merging \acp{BBH} coming from each mass transfer type.}
    \label{fig:m1-p}
\end{figure*}

\begin{figure*}
    \includegraphics[width=\textwidth]{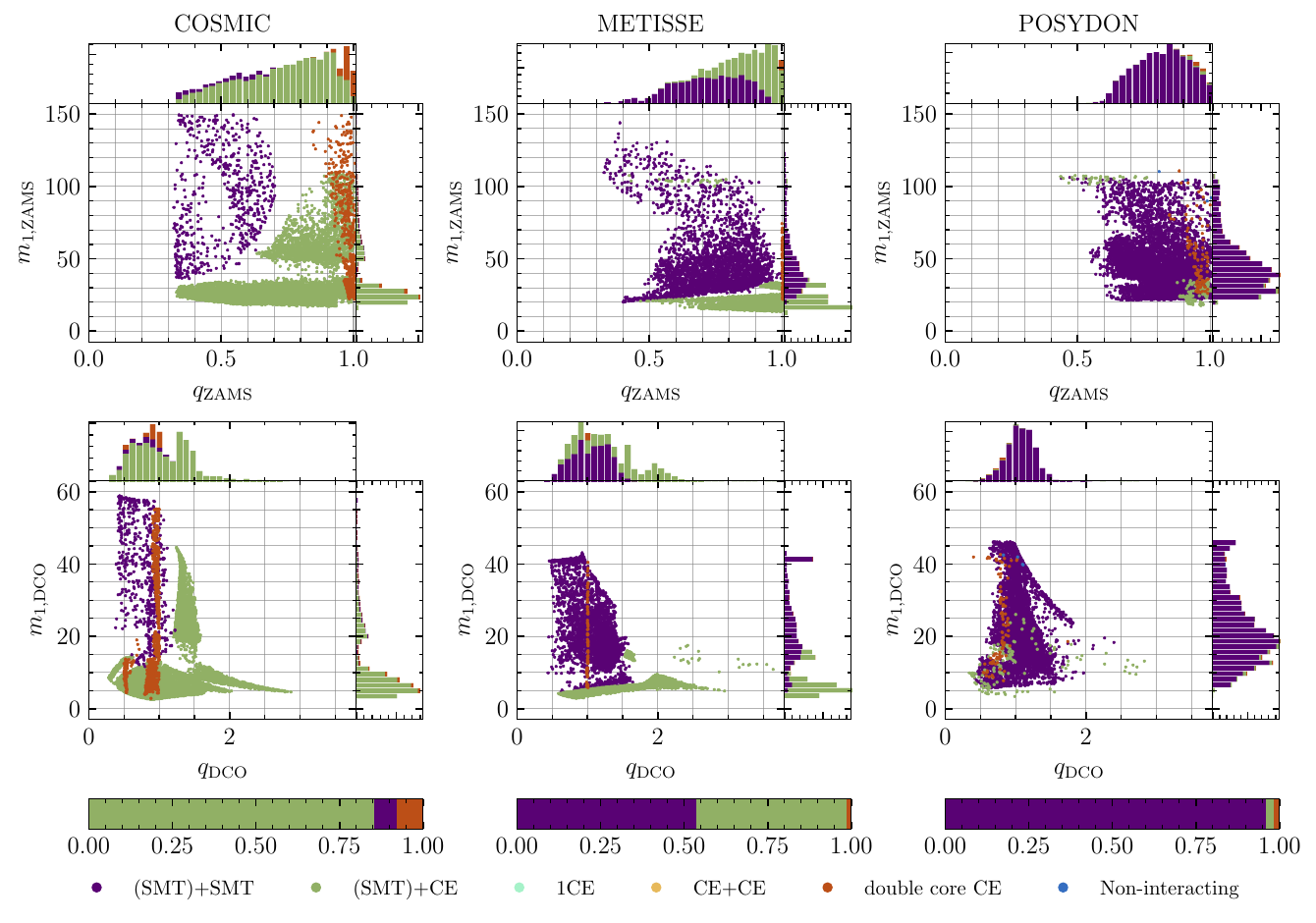}
    \caption{Same as Figure~\ref{fig:m1-p} but for mass ratio and primary mass.}
    \label{fig:q-m1}
\end{figure*}

\begin{figure*}
    \includegraphics[width=\textwidth]{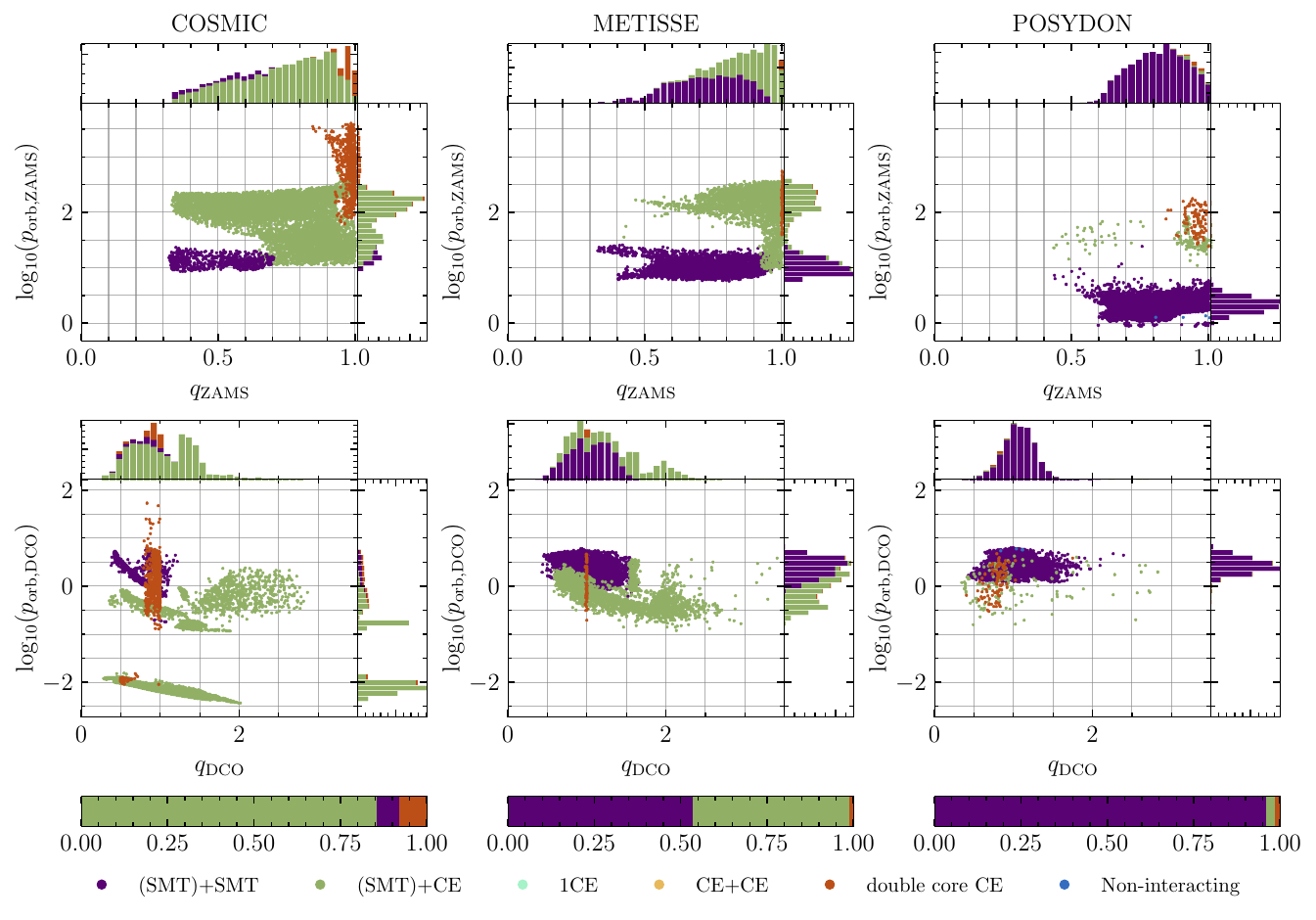}
    \caption{Same as Figure~\ref{fig:m1-p} but for mass ratio and orbital period.}
     \label{fig:q-p}
\end{figure*}
After generating the initial $0.01 Z_\odot$ binary population, the binaries are evolved with \COS{}, \MET{}, and \POS{}. We then select only those systems that produce \acp{BBH} that merge within a Hubble time. In Figures~\ref{fig:m1-p},~\ref{fig:q-m1}, and~\ref{fig:q-p}, we investigate the two-dimensional and marginal distributions for $(m_1,p_{\rm orb})$, $(q,m_1)$, and $(q,p_{\rm orb})$, respectively. Top panels depict the \ac{ZAMS} properties whereas the bottom panels depict these properties at the moment of \ac{BBH} formation. The scatter points in the two-dimensional distributions each indicate a single system in our population that became a \ac{BBH} merger in the code it was evolved with. We also indicate the mass transfer phases each system went through and the total percent of systems that underwent each mass transfer scenario. Note that $m_1$ always depicts the mass of the initially more massive star at \ac{ZAMS}. Therefore, $m_{1, \rm DCO}$ is the mass of the black hole sourced by the initially more massive star at \ac{ZAMS}. This differs from the primary mass as defined in the \ac{GW} community, which defines the primary as the more massive black hole at \ac{DCO} formation. We also stress that Figures~\ref{fig:m1-p},~\ref{fig:q-m1}, and~\ref{fig:q-p} are populations from a single metallicity, and have not been convolved with a star formation history or metallicity distribution. Therefore, the marginal distributions shown here are not directly comparable to the \ac{BBH} distributions found by the LVK collaboration~\citep{GWTC5_catalog,GWTC5_pop}. 

The progenitor and final properties of merging \acp{BBH} deviate significantly between the three codes. In Figure~\ref{fig:m1-p}, we see that there is a typical lower bound in the \ac{ZAMS} orbital period across all codes. However, the lower bound is at much smaller initial orbital periods in \POS{} as opposed to \COS{} and \MET{}. The \ac{ZAMS} orbital period for systems in \COS{} and \MET{} have lower limits of around $\sim 10$ days and $5.6$ days, respectively. On the other hand, \POS{} has a lower limit of $\sim 1$ day~\citep[consistent with][]{Briel26}. The lower limits on $p_{\rm orb, ZAMS}$ stem directly from the stability of mass transfer. Binaries that begin at very close orbital periods will typically undergo a contact phase and begin unstable mass transfer. However, these systems will not have enough orbital energy to successfully eject their \acp{CE}, and will merge as stars. Since the stellar physics is roughly equivalent between \MET{} and \POS{}, the main differences between these two codes must arise from differences in the binary physics. \POS{} binaries can become \ac{BBH} mergers when beginning at lower \ac{ZAMS} orbital periods due to both the differences in the prescription governing the onset of \ac{CE}, as well as having the ability to evolve the binary physics concurrently with the stellar physics. As will be explored more in Section~\ref{sec:POS-SMT}, binaries that are initially at very small \ac{ZAMS} orbital periods typically begin Case A mass transfer~\citep[when the donor is on the main-sequence, see also][]{Briel26}. In \COS{} and \MET{}, these initially close systems' orbits decay very rapidly, leading to stellar mergers. In \POS{}, however, the orbital period decays on a much slower timescale, allowing the primary to transfer enough mass to its companion for a mass ratio reversal to occur. This allows the orbit to widen, preventing a stellar merger. These \ac{BBH} mergers going through the (SMT)$+$SMT channel (where SMT stands for stable mass transfer and (SMT)$+$SMT indicates a binary going through no or stable mass transfer before the first \ac{SN} and solely stable mass transfer after the first \ac{SN}) highlight a key region of \ac{ZAMS} parameter space that drastically changes when incorporating more realistic binary physics. Note that there are also 4 binaries in \POS{} at small \ac{ZAMS} orbital periods that merge in the ``Non-interacting" channel. These binaries are likely undergoing chemically homogeneous evolution and stay at small enough radii to avoid any mass transfer phases before \ac{DCO} formation.

As seen in the bar graphs in the bottom of Figure~\ref{fig:m1-p}, \COS{} systems are much more likely to go through a \ac{CE} event than \MET{} or \POS{}. Specifically, $\sim85\%$ of \COS{} systems go through the (SMT)$+$CE channel, with another $\sim 8\%$ of systems going through the double core \ac{CE} channel. On the other hand, $\sim 54\%$ of merging \acp{BBH} in \MET{} and $\sim95\%$ of merging \acp{BBH} in \POS{} go through the (SMT)$+$SMT channel. The higher proportion of (SMT)$+$SMT systems in \MET{} than \COS{} follows from discrepancies in the radial evolution affecting the time the first stable mass transfer phase begins as well as the amount of orbital widening a binary may experience if a mass-ratio reversal occurs (see e.g. Figure~\ref{fig:VDH_COS-SMTCE_MET-POS-none} and the discussion in Section~\ref{sec:indiv-bins}). The additional proportion of (SMT)$+$SMT systems in \POS{} is due to the self-consistent modeling of stellar evolution---including angular momentum transport inside the stars, between them, and the orbit---with binary physics, which can affect the radial evolution and orbital dynamics of closely-separated systems.

The differences in the marginal $m_{1, \rm ZAMS}$ distributions in Figure~\ref{fig:m1-p} come from the mass transfer physics, although, the upper and lower bounds are from the \ac{SN} prescriptions. The upper limit on \ac{ZAMS} masses is set by the \ac{PISN} prescription. However, as seen in Figure~\ref{fig:mrem}, there are binaries that transfer enough mass to avoid a \ac{PISN}. The lower limit on \ac{ZAMS} masses is set by the remnant mass prescription, as primaries with low enough \ac{ZAMS} masses will form neutron stars instead of black holes.

The top panels of Figure~\ref{fig:q-m1} show the mass ratio distributions for \ac{BBH} progenitors. The structure, again, is quite different depending on which code is used. Progenitors that undergo double core \ac{CE} are peaked strongly at $q\sim 1$ by definition. This is because stars with near-equal mass ratios evolve similarly, and typically overflow their Roche lobes at the same time. \COS{} can form \acp{BBH} with \ac{ZAMS} mass ratios as low as $q \sim 0.3$, whereas \MET{} has a lower limit of $q \sim 0.35$, and \POS{} has a lower limit of $q \sim 0.5$. In \POS{}, binaries that have very unequal masses tend to go through a failed \ac{CE} phase, and merge as stars.

The differences in the initial \ac{ZAMS} populations that become merging \acp{BBH}, along with their mass transfer histories, lead to major differences in the final properties of merging \ac{BBH}s as well (see the bottom panels of Figures~\ref{fig:m1-p}, \ref{fig:q-m1}, and~\ref{fig:q-p}). For example, while \MET{} and \POS{}  produce the \ac{PISN} cutoff around $45 M_\odot$, the difference in how \COS{} treats \ac{PISN} systems allows \acp{BBH} to have larger $m_{1, \rm DCO}$ (see primary mass marginal distributions in Figures~\ref{fig:m1-p} and~\ref{fig:q-m1}). More generally, the shapes of the primary mass distributions are very different across codes. \COS{} yields two larger peaks, one broad peak around $\sim 4\mbox{--}12 M_\odot$ and a smaller peak around $21 M_\odot$. \MET{} has a sharper peak around $6 M_\odot$ mostly due to \ac{CE} evolution, and a secondary peak around $15 M_\odot$. On the other hand, \POS{} has two broad, overlapping peaks around $12 M_\odot$ and $18 M_\odot$.

One important consequence of defining the primary black hole from the initially more massive star at \ac{ZAMS} is that we can directly probe the amount of mass ratio reversal seen in each code. Since the mass ratio is defined as $q \equiv m_2/m_1$, in cases where the initially less massive star at \ac{ZAMS} becomes the more massive black hole, $q_{\rm DCO} > 1$. We see in Figures~\ref{fig:q-m1} and~\ref{fig:q-p} that, although \CosmicMRRPerc of binaries in \COS{} undergo mass ratio reversal, \PosydonMRRPerc  of \POS{} systems that merge as \acp{BBH} undergo mass-ratio reversal, and \MetisseMRRPerc of systems in \MET{} undergo mass-ratio reversal. Mass-ratio reversal has been found to occur when there is an efficient stable mass transfer phase before the first \ac{SN}~\citep{2022ApJ...933...86Z,2022ApJ...938...45B,2026arXiv260521580S,Chen2026,Maclean26}. To determine where the majority of mass-ratio reversal systems originate, we also investigate the specific case of stable mass transfer these progenitors experience before the first \ac{SN}. We define Case A mass transfer as occuring when the donor is a MS star at the onset of \ac{RLOF}, Case B mass transfer when the donor is a post-MS star, and Case C mass transfer when the donor has already begun core He burning at the onset of \ac{RLOF}. $61 \%$ of (SMT)$+$SMT systems experience Case A mass transfer and $39 \%$ experience Case C mass transfer in \COS{}, but only $6 \%$ and $18 \%$ of these experience mass-ratio reversal, respectively. On the other hand, $\sim 20 \%$ and $\sim 80 \%$ of (SMT)$+$CE systems go through Case B and Case C mass transfer in \COS{}, respectively. $100 \%$ and $\sim 41 \%$ of these systems experience mass-ratio reversal. In \MET{},  $\sim 87 \%$ of (SMT)$+$SMT systems go through Case A mass transfer before the first \ac{SN} and $\sim 99 \%$ of (SMT)$+$CE systems go through Case B mass transfer, with $\sim 50 \%$ and $\sim 74 \%$ of these systems undergoing mass-ratio reversal, respectively. Finally, over $97 \%$ of \ac{BBH} mergers in \POS{} that experience a stable mass transfer phase before the first \ac{SN} go through Case A mass transfer. Of these, $\sim 64 \%$ experience mass-ratio reversal. Across all codes, Case B stable mass transfer is the most efficient at producing mass-ratio reversal in \ac{BBH} progenitors. The reason for this stems from the mass ratio of the progenitor system at the start of \ac{RLOF}. The systems in all three codes that begin Case B stable mass transfer during the first mass transfer phase are more likely to have near-equal mass ratios at the start of \ac{RLOF}, leading to a higher mass transfer efficiency. 

Finally, the orbital period at the time of \ac{BBH} formation, $p_{\rm orb, DCO}$, has an upper limit around $5.6$ days in all codes, since binaries with orbital periods larger than this will not merge in a Hubble time.\footnote{Note that there are a few systems that have a higher DCO orbital period than this limit in \COS{} due to these systems having a non-zero eccentricity at \ac{BBH} formation. Although we do not consider \ac{SN} kicks, there is still non-zero eccentricity after a \ac{SN} due to the Blaauw mass-loss kick~\citep{Blaauw1961}.} \POS{} \acp{BBH} have orbital periods that are tightly clustered around $\sim 3.2$ days, similar to the \ac{ZAMS} orbital period distribution in \POS{}. This lower orbital period limit is due to the balance of needing an unequal mass ratio during stable mass transfer to efficiently shrink the orbit while being fundamentally limited by the onset of unstable mass transfer~\citep{Briel26,2026A&A...706A.296K}. \MET{} likewise has the same cluster around $3.2$ days as \POS{} for those binaries that go through the (SMT)$+$SMT channel, but also has an additional contribution of binaries with lower orbital periods at \ac{BBH} formation from the (SMT)$+$CE channel, as these binaries' orbits tighten more during unstable mass transfer. \COS{} (SMT)$+$SMT systems are again clustered around $\sim 3.2$ day orbital periods at \ac{BBH} formation. However, there is a large cluster of systems in the (SMT)$+$CE channel in \COS{} that have orbital periods around $0.008$ days at the time of \ac{BBH} formation. These low-period systems typically go through two \ac{CE} phases after the first \ac{SN}: the first \ac{CE} phase typically occurs when the secondary is core He-burning and the second \ac{CE} phase typically occurs when the secondary evolves off the helium main sequence as a stripped He star.

Note that $p_{\rm orb, DCO}$ (along with eccentricity) can be translated into a delay time distribution. Although we postpone discussion on delay time distributions to Section~\ref{sec:discussion}, we can see directly from the orbital period distributions that the three codes will have differing delay time distributions. Beyond the marginal distributions discussed above, we also see that the two-dimensional progenitor and \ac{DCO} distributions are likewise very different between the three codes, and all display different correlations between parameters. Therefore, it will be challenging to learn about the stellar progenitors of merging \acp{BBH} without some prior knowledge of the accuracy and validity of a given \ac{BPS} code.

\subsection{Systems rarely become BBH mergers across multiple codes\label{sec:overlaps}}

As seen in Section~\ref{sec:scatters}, an initial population of stellar binaries can evolve into vastly different \ac{BBH} populations depending on which \ac{BPS} code is used, even when the evolutionary prescriptions are chosen to be as consistent as possible. From here, we now investigate whether there are any robust regions of parameter space where the same systems will lead to \ac{BBH} mergers regardless of which \ac{BPS} code is used.

\begin{figure}
    \includegraphics[width=\linewidth]{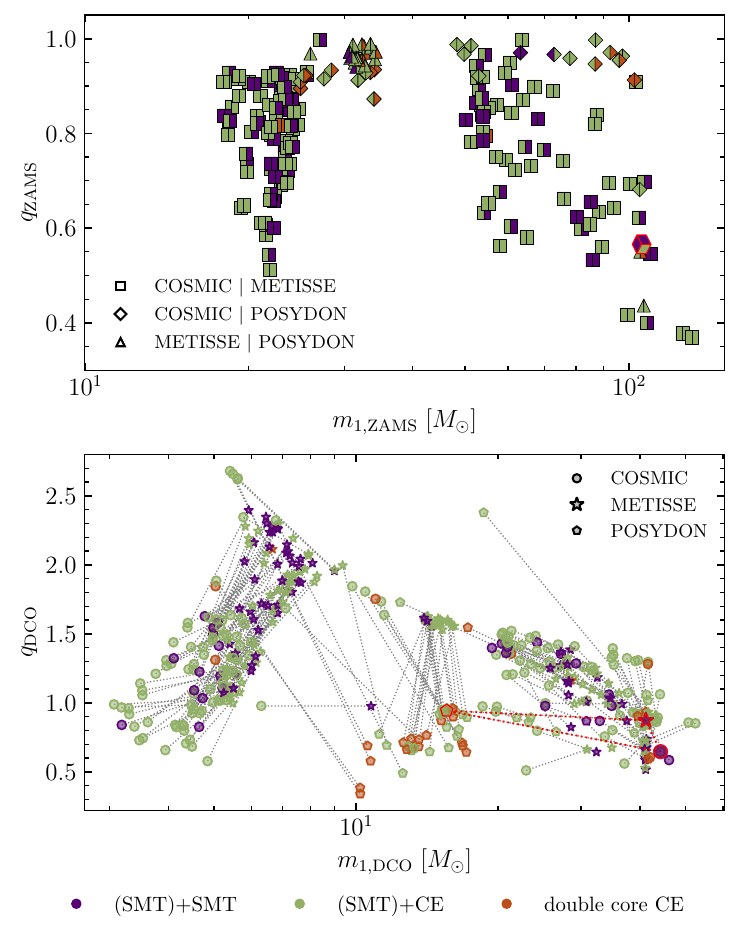}
    \caption{(Top panel) Individual binaries that merge as \ac{BBH}s in both \COS{} and \MET{} (\CosmicMetisseOverlapN; squares), \COS{} and \POS{} (\CosmicPosydonOverlapN; diamonds), or \MET{} and \POS{} (\MetissePosydonOverlapN; triangles), for a fiducial metallicity population of  $Z = 0.01 Z_\odot$. In both the top and bottom panels, we down-sample and plot a random subset of 150 out of the \CosmicMetisseOverlapN binaries that merge in \COS{} and \MET{} for clarity. The colors of the data points indicate what mass transfer history each binary goes through for \mbox{(\texttt{code 1} $|$ \texttt{code 2})}. There is one individual binary that leads to a \ac{BBH} merger in all three codes (hexagon outlined in red and colored according to the mass transfer history in \COS{} on the left, in \MET{} on the top right, and in \POS{} on the bottom right). (Bottom panel) Same binaries as the top panel, but now showing the resulting \ac{DCO} properties for \texttt{code 1} and \texttt{code 2} separately, with a gray dotted line connecting each individual binary. The binary that merges in all three codes is outlined in red and connected by red dotted lines. The disconnect between \ac{DCO} parameters in each code demonstrates that even when the same individual binary becomes a merging \ac{BBH} in two codes, the final \ac{DCO} parameters are typically dissimilar.}
    \label{fig:same-bins}
\end{figure}

In our initial ${Z=0.01 Z_\odot}$, $0.1 Z_\odot$, and $1.0 Z_\odot$ populations, with $10^6$ binaries each, we find \ThreeCodeOverlap systems that merge as \acp{BBH} in all three \ac{BPS} codes, respectively. However, there are more systems that become \ac{BBH} mergers in two out of the three codes. In Figure~\ref{fig:same-bins}, we take the initial binaries in our fiducial $0.01 Z_\odot$ population and plot each binary that becomes a \ac{BBH} merger in both \COS{} and \MET{} (squares),\footnote{In Figure~\ref{fig:same-bins}, we plot only a random subset of 150 mergers out of the \CosmicMetisseOverlapN that are consistent between \COS{} and \MET{} for visual clarity.} \COS{} and \POS{} (diamonds), or \MET{} and \POS{} (triangles). Each binary that is found to merge in two of the three codes is colored on the left and right based on the mass transfer history that led to the \ac{BBH} in that system in each code. We additionally plot the single system that merges in all three codes with a red outline.

There are a couple regions of parameter space where matching \ac{BBH} systems go through similar mass transfer scenarios. For example, many binaries that merge as \acp{BBH} in both \COS{} and \MET{} with $m_{1, \rm ZAMS} \sim 20 M_\odot$ go through the (SMT)$+$CE channel and systems with with $m_{1, \rm ZAMS} \sim 100 M_\odot$ both go through the (SMT)$+$SMT channel. However, there are also systems that merge in multiple codes that go through different mass transfer histories. For example, a binary that leads to a merging \ac{BBH} with $m_{1, \rm ZAMS} \sim 105 M_\odot$ and $q_{\rm ZAMS} \sim 0.44$ goes through the (SMT)$+$SMT channel in \MET{} but the (SMT)$+$CE channel in \POS{}.

 Out of the $10^6$ binaries in our $0.01 Z_\odot$ population, \CosmicMetisseOverlapN become \ac{BBH} mergers in \COS{} and \MET{}, \CosmicPosydonOverlapN become \ac{BBH} mergers in \COS{} and \POS{}, and \MetissePosydonOverlapN become \ac{BBH} mergers in both \MET{} and \POS{}. Another way to assess the amount of overlap between the \ac{BBH} merger populations found in the three codes is using the Dice-S{\o}renson coefficient~\citep[DSC;][]{Dice,sorensen1948method}, defined as the overlap between two sets divided by the average length of two sets. \COS{} finds \CosmicBBHMergers \ac{BBH} mergers in the $0.01 Z_\odot$ population, \MET{} finds \MetisseBBHMergers, and \POS{} finds \PosydonBBHMergers. Therefore, \COS{} and \MET{} have a DSC of \CosmicMetisseDSC, or a \CosmicMetisseOverlapPerc overlap. \COS{} and \POS{} have a \CosmicPosydonOverlapPerc overlap, and \MET{} and \POS{} have a \MetissePosydonOverlapPerc overlap. 
 
 While the top panel of Figure~\ref{fig:same-bins} does demonstrate that there are regions of parameter space where two codes become more consistent with each other (for example, the cluster of squares around an $m_{1, \rm ZAMS} \sim 20 M_\odot$), there is a $< 15\%$ overlap in systems that become \ac{BBH} mergers between any of the two \ac{BPS} codes across all three metallicity populations we consider. As such, we cannot confidently conclude that there are any regions of parameter space that are robust to differences in \ac{BPS} codes. Even with the near-identical stellar physics in \MET{} and \POS{}, the minimal overlap in binaries that produce \ac{BBH} mergers demonstrates that binary physics also plays a major role in the outcomes of compact object populations.

In the bottom panel of Figure~\ref{fig:same-bins}, we take each \ac{ZAMS} data point from the top panel and split it into two points for the final \ac{BBH} properties in \texttt{code 1} and \texttt{code 2}, connected by gray dotted lines and still colored based on the mass transfer history the binary goes through in each code. The single \ac{BBH} that merges in all three codes is connected by red dotted lines. It is evident that, even though there are binaries that become \ac{BBH} mergers in multiple codes, these binaries do not lead to consistent \ac{BBH} properties. For example, for systems that merge in both \MET{} and \POS{}, those that undergo mass-ratio reversal in \MET{} often do not in \POS{}, and vice versa, leading to different $q_{\rm DCO}$ even when the black hole masses are similar. Likewise, even when two codes both go through the same mass transfer scenario, the systems often lead to different black hole masses based on the two codes. \COS{} typically has systematically lower black hole masses when going through the (SMT)$+$CE channel and a core-collapse \ac{SN} than \MET{} and \POS{} for the same binary, demonstrating that \COS{} binaries seem to lose more mass during mass transfer. This is likely a result of \COS{} using the SSE tracks from~\citet{Hurley_SSE}, which predicts smaller stellar cores than in \texttt{MESA}. These differences in \ac{DCO} properties even when the \ac{BBH} is coming from the same system emphasizes the difficulties in using \ac{GW} populations to learn about \ac{BBH} progenitors.

\subsection{Investigating cross-code discrepancies in individual binaries\label{sec:indiv-bins}}
\begin{figure*}
    \centering
    \includegraphics[width=0.93\linewidth]{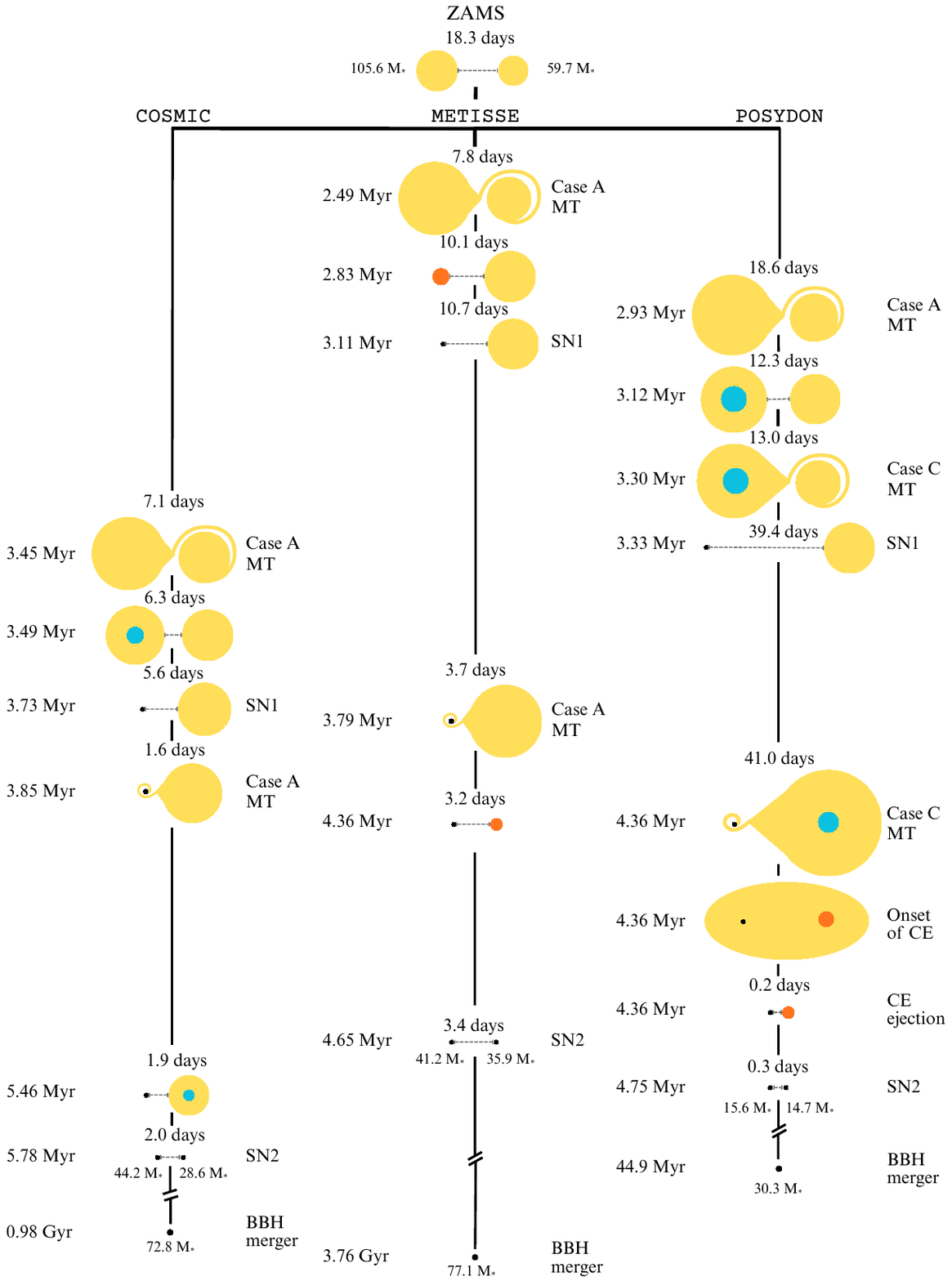}
    \caption{A branched Van den Heuvel diagram for the single system that consistently becomes a \ac{BBH} merger in \COS{} (left branch), \MET{} (middle branch), and \POS{} (right branch). Time is depicted along the y-axis, but is not drawn to scale. Evolution times are provided on the left, important phases in the evolution are provided on the right, and orbital periods are provided above each step. Yellow circles indicate MS stars. Yellow circles with orange centers are stars on the Hertzsprung gap. Yellow circles with blue centers are stars that have begun core He burning. Bare orange circles are stripped He stars. Black circles indicate a black hole.}
    \label{fig:VDH_COS-MET-POS}
\end{figure*}

While some systems with identical \ac{ZAMS} properties consistently produce \ac{BBH} mergers across multiple codes, the majority do not. Furthermore, only one system in the $0.01 Z_\odot$ population becomes a \ac{BBH} merger in all three codes. To diagnose the mismatches found above, we now directly compare key evolutionary phases of three example binaries across all three codes to pinpoint mechanisms driving these divergences. We first examine the lone consistent \ac{BBH} merger to determine how this consistency is achieved despite the discrepancies in mass transfer history across codes. We then analyze two additional binaries in our $0.01 Z_\odot$ population that do not align between the \ac{BPS} codes and occupy interesting regions of parameter space.

\begin{figure*}
    \centering
    \includegraphics[width=\linewidth]{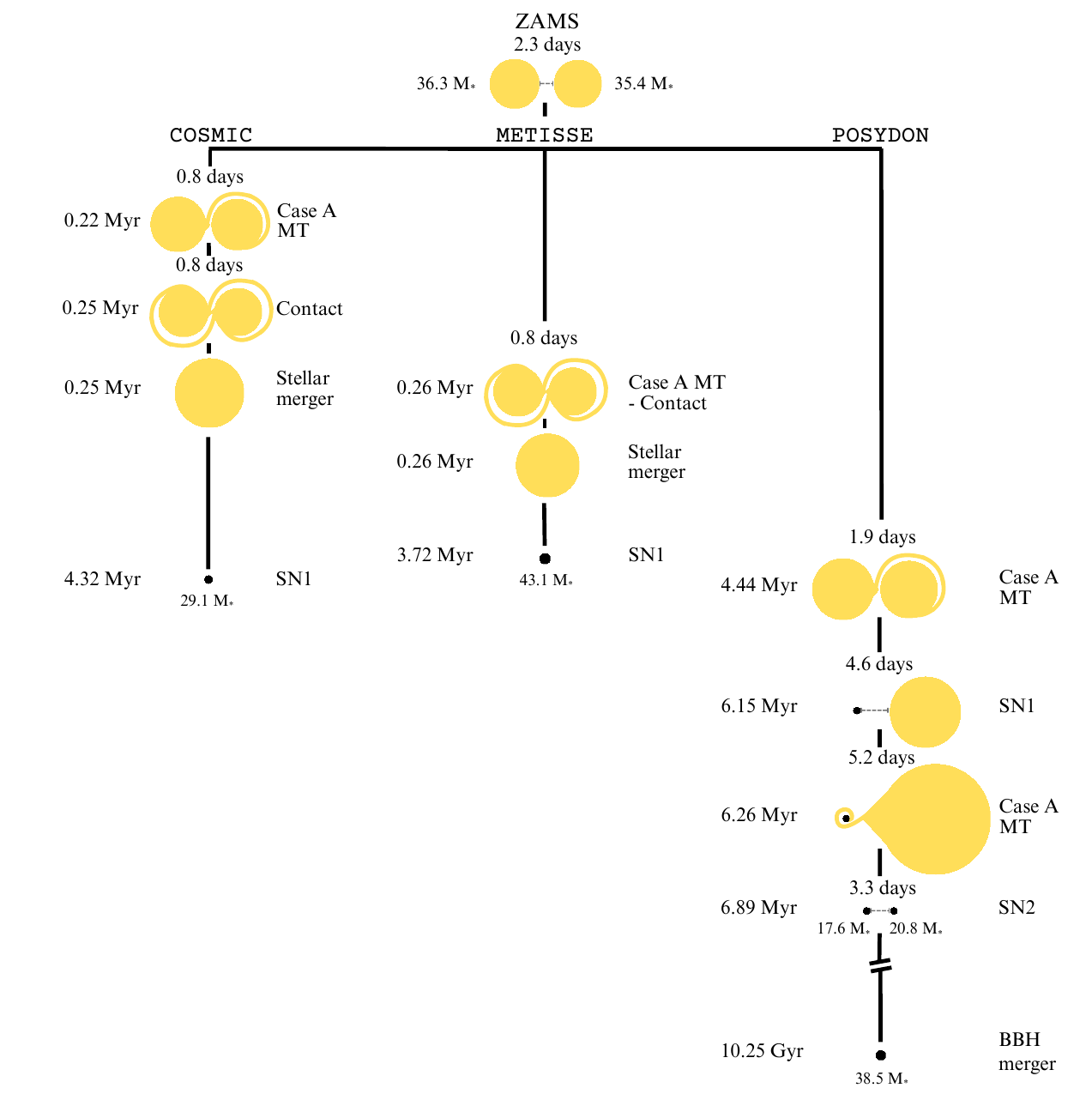}
    \caption{Same as Figure~\ref{fig:VDH_COS-MET-POS}, but for an individual binary that merges as a \ac{BBH} in \POS{} (right branch) through the (SMT)$+$SMT channel, but does not become a merging \ac{BBH} in \COS{} (left branch) or \MET{} (middle branch).}
    \label{fig:VDH-POS-SMT_MET-COS-none}
\end{figure*}

\begin{figure*}
    \centering
    \includegraphics[width=0.938\linewidth]{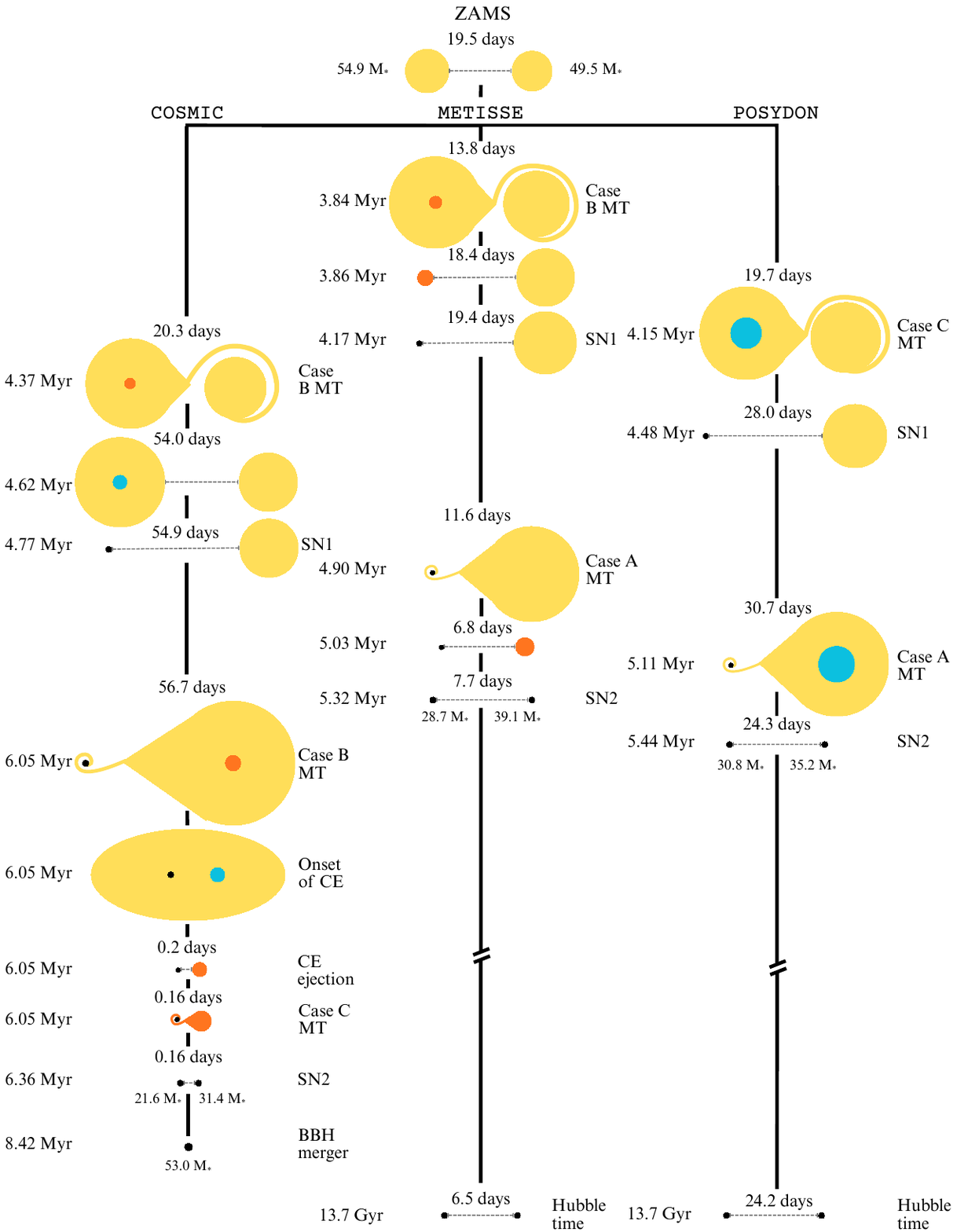}
    \caption{Same as Figure~\ref{fig:VDH_COS-MET-POS}, but now for an individual binary that becomes a \ac{BBH} merger in \COS{} through the (SMT)$+$CE channel (left branch), but does not merge in \MET{} (middle branch) or \POS{} (right branch).}
    \label{fig:VDH_COS-SMTCE_MET-POS-none}
\end{figure*}

\subsubsection{The consistent BBH merger in all three codes\label{sec:3-consistent}}

The system that becomes a \ac{BBH} merger in all three codes has a \ac{ZAMS} orbital period of $18.3$ days, with $m_{1, \rm ZAMS} = 105.6 M_\odot$ and  $q_{\rm ZAMS} = 0.57$. As seen in Figure~\ref{fig:same-bins}, this system goes through the (SMT)$+$SMT channel in both \COS{} and \MET{}, but goes through the (SMT)$+$CE channel in \POS{}, and the final \ac{BBH} has different properties across all codes. To further investigate the evolutionary history of this binary, we schematically draw important evolutionary steps that occur in each code using a Van den Heuvel (VdH) diagram~\citep{1975A&A....39...61F,1981SSRv...30..623V} in Figure~\ref{fig:VDH_COS-MET-POS}, with the left branch depicting the binary's evolution when using \COS{}, the middle branch when using \MET{}, and the right branch when using \POS{}. 

In all three codes, the first phase of mass transfer is Case A stable mass transfer. \MET{} and \POS{} stars expand more on the main sequence than stars in \COS{}, leading these stars to overflow their Roche lobes earlier in their evolution. \COS{} and \MET{} both experience a mass-ratio reversal during this first stable mass transfer phase, although the mass-ratio reversal occurs much earlier in \MET{}. The earlier onset and prolonged duration of the mass-ratio reversal phase in \MET{} leads to an orbital widening during this first mass transfer phase compared to the same binary in \COS{}. When the binary is instead evolved with \POS{}, the donor goes through two stable mass transfer events, with a brief detached phase when the donor is core He-burning. The accretion efficiency in \POS{} is much lower due to the rapid spin-up of the accretor compared to the mostly conservative mass transfer in \COS{} and \MET{}, and as such the binary does not experience a mass-ratio reversal event in \POS{}. In \COS{} and \MET{}, the donor stops mass transfer before core-collapse, when the donor is a core He-burning star and a stripped He star, respectively. The donor in \POS{} continues transferring mass until core carbon-depletion, where it then undergoes core-collapse. The mass of the primary black hole directly after \ac{SN} is $42.3 M_\odot$ in \COS{}, $40.5 M_\odot$ in \MET{}, and $15.6 M_\odot$ in \POS{}. 

The second phase of mass transfer is where \POS{} differs the most from \COS{} and \MET{}. The secondary in \COS{} and \MET{} again starts Case A stable mass transfer onto the companion black hole. At the start of \ac{RLOF}, the mass ratio of the donor over the accretor is $2.67$ in \COS{} and $2.43$ in \MET{}. These mass ratios are less than the prescribed $q_{\rm crit}$ for a MS star, and thus the mass transfer remains stable. However, directly after the first \ac{SN}, the \POS{} binary is at a much larger orbital period, and does not begin mass transfer until the secondary is core He-burning. Note that the larger orbital period is exclusively from differences in binary physics since we do not consider \ac{SN} kicks here. At this point, the secondary's radius is so large that it triggers a \ac{CE} phase in \POS{}, leading to rapid orbital hardening with the ejection of the envelope. In \COS{} and \MET{}, another mass-ratio reversal event occurs during the second stable mass transfer phase, ultimately preserving the overall mass-ratio hierarchy in this system. The final orbital period after the second \ac{SN} in \POS{} is much shorter due to the \ac{CE} phase, and thus the resulting \ac{BBH} in \POS{} merges on a much faster timescale, with a delay time of $44.9$ Myr as opposed to a $0.98$ Gyr and $3.76$ Gyr delay time in \COS{} and \MET{}, respectively.

\subsubsection{BBH merger in \POS{} that goes through solely stable mass transfer\label{sec:POS-SMT}}

The vast majority of systems that become \ac{BBH} mergers in \POS{} are (SMT)$+$SMT systems that have an initial orbital period of $\lesssim 3.2$ days (see the cluster of purple points at low \ac{ZAMS} orbital periods in the top right panel of Figure~\ref{fig:m1-p}). However, no systems with $p_{\rm orb, ZAMS} \lesssim 5.6$ days become \ac{BBH} in \COS{} or \MET{}. To investigate why \POS{} systems are able to yield successful \ac{BBH} mergers at such close periods, we now examine a binary that has a \ac{ZAMS} orbital period of 2.3 days, with $m_{1, \rm ZAMS} = 36.3 M_\odot$ and  $q_{\rm ZAMS} = 0.98$. 

Figure~\ref{fig:VDH-POS-SMT_MET-COS-none} depicts a branched VdH diagram of this binary. The primary begins Case A stable mass transfer in all three codes, although in \MET{}, both the primary and secondary overfill their Roche lobes, leading to a contact system. When evolving the system with \COS{} or \MET{}, the binary orbital period decays quickly. When stable mass transfer begins in these codes, the stars swiftly come into contact and merge due to the extremely small initial orbital period. When using \POS{} to evolve the binary, the orbital period does not decay as rapidly. The difference likely originates from the treatment of stellar rotation. Where in \POS{} the models are initially synchronized with the orbit, \COS{} and \MET{} follow BSE in setting the initial rotation, which is not synchronized with the orbit initially~\citep{Hurley_SSE,Hurley_BSE}. As such, angular momentum is redistributed from the orbit to the stars in \COS{} and \MET{}, causing the rapid orbital shrinkage in these codes. In \POS{}, stars in initially tight orbits spin quite rapidly, leading to rotational mixing~\citep{2009A&A...497..243D,2024IAUS..361..343H}, which alters the stellar structure compared to a non-rotating model and remains more compact, altering the mass transfer stability. The self-consistent treatment of stellar and binary physics in \POS{} captures the stellar rotation's impact on the orbit and stellar structure. 

Due to the differences in stellar rotation, Case A stable mass transfer begins much later in \POS{}, but at a slightly wider orbital period. As Case A stable mass transfer continues in \POS{}, the binary experiences mass-ratio reversal; the binary orbit widens enough that the binary avoids merging, and the primary eventually collapses into a black hole. From here, the secondary also begins a stable Case A mass transfer phase onto the black hole until it also collapses. The final black hole masses in \POS{} are $m_{1, \rm DCO}=17. 6 M_\odot$ and $m_{2, \rm DCO} = 20.8 M_\odot$, preserving the mass-ratio reversal. The \ac{BBH} inspirals until it merges with a delay time of 10.25~Gyr. The evolutionary history of this system and the rest of the (SMT)$+$SMT channel systems in \POS{} are broadly consistent with stable mass transfer \ac{BBH} merger populations found in other works using detailed \ac{BPS} codes~\citep[see e.g.][]{Briel26}.

\subsubsection{BBH merger in \COS{} that goes through a \ac{CE} phase\label{sec:COS-CE}}

The majority of systems in \COS{}, on the other hand, merge through the (SMT)$+$CE channel. While there are systems evolved with \MET{} or \POS{} that also yield \ac{BBH} mergers in the same regions of parameter space as \COS{}, there are many systems in \COS{} with $40 M_\odot \lesssim m_{1, \rm ZAMS} \lesssim 100 M_\odot$ that become \ac{BBH} mergers that do not when instead evolved with \MET{} and \POS{} (see green points in the top left panel of Figure~\ref{fig:m1-p}). Here, we investigate one of these systems that merges through the (SMT)$+$CE channel in \COS{}, but does not produce a \ac{BBH} merger in \MET{} or \POS{}. This binary has a \ac{ZAMS} orbital period of 19.5 days, with $m_{1, \rm ZAMS} = 54.9 M_\odot$ and  $q_{ \rm ZAMS} = 0.90$. To illustrate the key evolutionary phases this binary experiences when evolved with each code, we again construct a branched VdH diagram in Figure~\ref{fig:VDH_COS-SMTCE_MET-POS-none}. 

As seen in Figure~\ref{fig:HRdiagram}, for a $\sim 50 M_\odot$ single star, \POS{} begins core He-burning much sooner in the star's evolution. Therefore, while the donor in Figure~\ref{fig:VDH_COS-SMTCE_MET-POS-none} begins Case B stable mass transfer in \COS{} and \MET{}, the donor goes through Case C stable mass transfer in \POS{}. HG stars in \MET{} also expand more rapidly, and thus the donor in \MET{} overflows its Roche lobe much earlier than in \COS{}. During the first stable mass transfer phase, all three codes experience a mass-ratio reversal between the donor and accretor. However, while \MET{} only experiences mass-ratio reversal once the donor's envelope is stripped, the donor in \COS{} and \POS{} experiences the mass-ratio reversal earlier, and the orbital period widens significantly, especially in \COS{}. While \COS{} and \MET{} end their mass transfer phases before the donor collapses, the donor in \POS{} transfers mass to its companion all the way until it experiences core-collapse. The mass of the primary black hole directly after \ac{SN} is $21.6 M_\odot$ in \COS{}, $28.6 M_\odot$ in \MET{}, and $30.5 M_\odot$ in \POS{}. 

After the first \ac{SN}, the binary remains at a tighter orbital period in \MET{}. Once the secondary expands, it begins \ac{RLOF} while on the main sequence, and begins Case A stable mass transfer. Due to the larger orbital separations in \COS{} and \POS{}, the secondary begins \ac{RLOF} post-MS. As with the primary, the secondary in \POS{} begins core He-burning much quicker in its evolution, and thus the secondary begins Case C stable mass transfer, as opposed to Case B stable mass transfer in \COS{}. At the start of \ac{RLOF} in \COS{}, $m_{\rm donor} / m_{\rm acc} = 3.6$. While this is below the critical mass ratio for HG stars, when the secondary begins core He-burning in \COS{}, this mass ratio is now larger than the critical mass ratio $q_{\rm crit} = 3$ for core He-burning stars, and a \ac{CE} phase begins. On the other hand, $m_{\rm donor} / m_{\rm acc} = 2.5$ in \MET{} at the start of the second mass transfer phase, and the mass transfer remains stable. Due to the difference in the onset of \ac{CE} in \POS{}, the second phase of mass transfer likewise remains stable. During the \ac{CE} phase in \COS{}, the binary experiences orbital hardening due to the ejection of the envelope, leading to an orbital period of $0.2$~days after the successful ejection. The stripped He star secondary begins a very brief Case C stable mass transfer phase with its companion black hole due to its small separation before collapsing to a black hole. Since the second mass transfer phase in \MET{} and \POS{} remains stable, the orbit does not shrink, and the orbital periods after the second \ac{SN} remain at $7.7$~days and $24.3$~days, respectively. The \ac{CE} phase in \COS{} brings the binary close enough to merge with a delay time of $8.42$~Myr. Meanwhile, the binary evolved with \MET{} or \POS{} does not merge within a Hubble time.

\section{Discussion\label{sec:discussion}}
We have found that different \ac{BPS} codes lead to large discrepancies in stellar properties, such as radius and remnant mass. These differences affect the mass transfer history for the binaries, and lead to major differences in the progenitor and merger properties of \acp{BBH}. 

The above results focus on specific single-metallicity populations to isolate the effects solely due to the choice of \ac{BPS} code. To translate these discrepancies in single-metallicity populations to differences in observable \ac{GW} populations, we would additionally need to convolve our single-metallicity populations with a star formation history (SFH) and metallicity distribution. However, the SFH and metallicity distribution also contain observational uncertainties~\citep{2019MNRAS.488.5300C,2021MNRAS.508.4994C,2024AnP...53600170C}, adding in another layer of uncertainty when comparing \ac{GW} populations to populations produced with various \ac{BPS} codes~\citep{2019MNRAS.490.3740N,2020MNRAS.493L...6T}. While we do not consider discrepancies in observed merger rates, we can still investigate the time delay distributions of \ac{BBH} mergers as a function of code and metallicity.

\begin{figure*}
    \includegraphics[width=\textwidth]{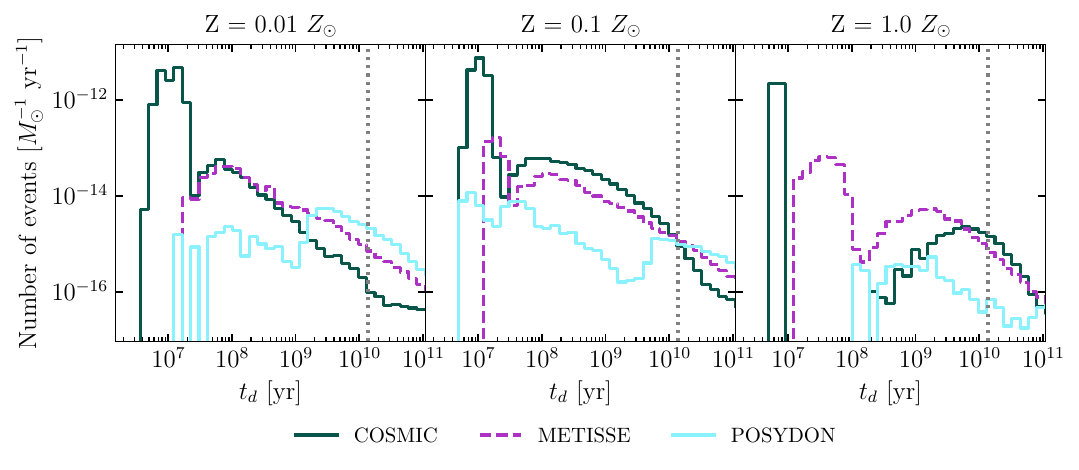}
    \caption{Delay time distribution of merging \acp{BBH} in \COS{} (dark green), \MET{} (magenta), and \POS{} (light blue) for 3 fiducial metallicities. The distributions are normalized to the total amount of mass in the population assuming a binary fraction of one. They gray dotted line denotes a Hubble time.}
    \label{fig:td}
\end{figure*}

In Figure~\ref{fig:td}, we show the distributions of delay times from our three fiducial metallicity populations, $0.01 Z_\odot$ (left panel), $0.1 Z_\odot$ (middle panel), and $1.0 Z_\odot$ (right panel) evolved with \COS{} (dark green), \MET{} (magenta), and \POS{} (light blue).\footnote{As mentioned previously, each population has been generated at a single epoch. Therefore, the delay times depicted in Figure~\ref{fig:td} are only for these single-age stellar populations.} Each delay time distribution is normalized by the total stellar mass in the population, assuming a binary fraction of one. As expected, the delay time distributions also show discrepancies between the three codes for all metallicities. In the $0.01 Z_\odot$ population, \COS{} is peaked at much shorter delay times due to the vast majority of \ac{BBH} mergers in \COS{} coming from the \ac{CE} channel, which efficiently tightens binary orbits. \POS{}, on the other hand, peaks at larger delay times due to the dominance of the (SMT)$+$SMT channel, which is less efficient in tightening binary orbits.\footnote{While not shown here, \POS{} is more efficient in merging \acp{BBH} when including \ac{SN} kicks, and including \ac{SN} kicks adds a contribution at shorter delay times to the delay time distribution~\citep[see e.g.][]{Briel26}. More generally, the delay time distributions for all three codes shown in Figure~\ref{fig:td} will change when including \ac{SN} kicks.} \MET{}, which has binaries going through both stable mass transfer and \ac{CE} channels, has a delay time distribution that peaks between the other two distributions. As metallicity increases, there is a larger contribution of \ac{BBH} mergers that go through a \ac{CE} event in \MET{} and \POS{} (see Appendix~\ref{sec:otherZ}), which increases the amount of mergers with shorter delay times. However, at solar metallicity, \ac{BBH} merger efficiencies are sufficiently low such that there are not enough mergers to produce a smooth delay time distribution. The differences in delay times found here will translate into differences in \ac{GW} merger rates. For example, shorter delay times lead to steeper \ac{BBH} merger redshift distributions.

In this work, we consider a fiducial set of stellar and binary prescriptions across the three \ac{BPS} codes. We have chosen the above prescriptions to maximize the similarity between codes, but other choices may be preferred. While we do not provide a detailed exploration of the effects different prescriptions have on our conclusions, it is possible that a different set of prescriptions will induce more or less similar final \ac{BBH} populations across the different \ac{BPS} codes. As an example of how different prescriptions can affect the mismatch between \ac{BPS} codes, we investigate two poorly-constrained prescriptions for the onset of \ac{CE} in \COS{}, \texttt{qcflag} and \texttt{smt$\_$periastron$\_$check}, in Appendix~\ref{sec:bin-variations}. These two prescriptions do not have an exact analog in detailed \ac{BPS} codes, but do have significant effects on the final BBH merger populations in rapid and hybrid \ac{BPS} codes. Since the choices we explore in Appendix~\ref{sec:bin-variations} are equally justifiable, our conclusions drawn in Section~\ref{sec:pops} can be substantially altered when different choices for \ac{CE} prescriptions are chosen. More specifically, while we attempt to find robust areas of similarity in \ac{BBH} merger parameter space in Section~\ref{sec:overlaps}, any consistent regions that are found across codes may not be robust for other evolutionary prescriptions. These variations highlight the necessity of accounting for model uncertainties both when comparing across \ac{BPS} codes and when using \ac{BPS} codes to interpret astrophysical populations. 

In Section~\ref{sec:pops}, we focused on discrepancies in the primary masses, mass ratios, and orbital periods of \acp{BBH} and their progenitors. However, there are additional parameters relevant to \ac{GW} sources, such as spin and eccentricity, that we do not consider in this work. We chose to omit spin due to the uncertainties in spin prescriptions and our inability to keep these prescriptions consistent between codes. Eccentricity, on the other hand, will typically be negligible since we do not consider \ac{SN} kicks and the binaries are assumed to circularize due to mass transfer. However, there will certainly also be differences in these parameters' distributions across \ac{BPS} codes generally. As these distributions will be highly variable due to other uncertainties, they warrant further investigation.

\ac{BPS} codes are used to investigate a myriad of astrophysical populations and phenomena. The differences in stellar and binary evolution found above will also lead to discrepancies in the populations of other astrophysical sources, such as neutron star-black hole and binary neutron star mergers, electromagnetic transients such as gamma-ray bursts, and Gaia populations. Other \ac{GW} formation channels, such as dynamical triples or star clusters, may also use \ac{BPS} codes as a dependency. Therefore, any inference on astrophysical populations using \ac{BPS} codes should carefully consider the inherent uncertainties that arise due to the choice of \ac{BPS} code.

\section{Conclusions\label{sec:conclusions}}
We have compared three \ac{BPS} codes: the rapid code \COS{}, the hybrid code \MET{}, and the detailed code \POS{}. We evolved a set of initial binary stars at three fiducial metallicities with each code, keeping the stellar and binary evolution prescriptions as consistent as possible outside of the inherent differences in the codes themselves. Our main results are summarized below:

\begin{itemize}
    \item Differences in stellar wind and core-envelope boundary prescriptions in \COS{} lead to deviations in single star evolution from the same stars evolved with \MET{} or \POS{}. Weaker winds in \COS{} lead to larger maximum radii for high-mass stars. Massive stars with \ac{ZAMS} masses below $\sim 130 M_\odot$ do not go through (pulsational) pair instabilities in \COS{} due to the use of SSE tracks, which predict smaller stellar cores than stars evolved with \MET{} and \POS{}. Smaller cores yield smaller black hole masses during core-collapse \ac{SN}, but larger black hole masses when \MET{} and \POS{} systems go through (P)\acp{PISN}. Incorporating binary interactions amplifies the existing divergences in stellar properties between the three codes.

    \item Stellar progenitors of \ac{BBH} mergers come from different parts of parameter space for each of the three codes due to the differences in stellar and binary physics between \COS{}, \MET{}, and \POS{}. In our fiducial $0.01 Z_\odot$ population, most progenitors that go through the (SMT)$+$SMT channel in all codes go through Case A stable mass transfer. Binaries evolved with \COS{} and \MET{} that begin at tight separations undergo rapid orbital decay, setting a lower limit on \ac{ZAMS} orbital periods. Binaries evolved with \POS{} can successfully create \ac{BBH} mergers at lower initial orbital periods due to the self-consistent implementation of stellar and binary physics. Differences in the prescriptions for the onset of \ac{CE} yield a much larger population of binaries that go through the (SMT)$+$CE channel in \MET{}, and even more so in \COS{}. $\sim 93 \%$ of systems in \COS{} go through a \ac{CE} channel, while $\sim 54 \%$ and $\sim 95 \%$ of systems in \MET{} and \POS{} experience a solely stable mass transfer scenario, respectively. 
    
    \item The disparate \ac{BBH} progenitor populations between \COS{}, \MET{}, and \POS{}, along with the treatment of binary physics, compound to major differences in single-metallicity \ac{BBH} merger populations. In the fiducial $0.01 Z_\odot$ population, \CosmicMRRPerc of \COS{} mergers experience mass-ratio reversal, as opposed to \MetisseMRRPerc and \PosydonMRRPerc in \MET{} and \POS{}, respectively.

    \item In the fiducial $0.01 Z_\odot$ population, there is \textit{one} binary that becomes a \ac{BBH} merger in all three codes. There is a $\sim 0-14 \%$ overlap in systems that become \ac{BBH} mergers between any two \ac{BPS} codes across all three of our fiducial metallicity populations, with less overlap at higher metallicities on average. Systems that lead to \ac{BBH} mergers in multiple codes show major differences in the \ac{BBH} properties at formation.

    \item \ac{BBH} mergers in \COS{} typically have much shorter delay times in the fiducial $0.01 Z_\odot$ population due to the majority of progenitors experiencing a \ac{CE} event. \ac{BBH} mergers in \POS{} have much longer delay times due to most \ac{BBH} progenitors going through solely stable mass transfer phases. The delay time distribution in \MET{} peaks between the two other codes. Variations in the delay time distributions will have a large effect on \ac{GW} populations, such as the \ac{BBH} redshift distribution. 
\end{itemize}

These results highlight the uncertainties in \ac{BPS} populations implicit in the choice of \ac{BPS} code. Specifically, the variation in \ac{BBH} merger populations from \COS{} to \MET{} demonstrate the sensitivity to varying stellar physics prescriptions. Likewise, the shift in the \ac{BBH} merger populations from \MET{} to \POS{} illustrates the impact of the implementation of binary physics. Recent works have also elucidated the inherent variation across \ac{BPS} codes in predicting \ac{GW} merger rates~\citep{2022LRR....25....1M,2026arXiv260605322B,2026arXiv260628515B}. \textit{The interpretation and use of \ac{BPS} outputs should be contingent on their contextualization within these inherent uncertainties.} The stark differences between these three \ac{BPS} codes from initially identical binary populations raise concerns regarding the physical uncertainties that apply to all \ac{BPS} models.

Future work should continue to investigate what regions of parameter space may be robust against code and prescription choices in order to minimize these potential systematics. However, improving \ac{BPS} codes' accuracy will also be paramount in minimizing the uncertainties that arise from using a given \ac{BPS} code. For example, our work has demonstrated that stellar winds and radii have a large impact on \ac{BBH} progenitor histories. Other recent work has also demonstrated the influence stellar winds have on black hole progenitors~\citep{2026arXiv260726147R}. Therefore, accurately modeling these prescriptions is essential.

Rapid, hybrid, and detailed \ac{BPS} codes inherently focus on different aspects of stellar and binary evolution, and each has its own benefits and drawbacks. To address some of the discrepancies we find here, the strengths of one \ac{BPS} code can be used to inform another. More specifically, the treatment of mass transfer physics in detailed codes can be approximated in rapid and hybrid codes to achieve a higher physical fidelity while preserving the ability to rapidly explore parameter space. As an example, \citealt{2026A&A...706A.296K} recently proposed a new treatment for mass transfer stability that rapid \ac{BPS} codes can implement that better aligns with detailed codes.

Finally, the vast wealth of observational data from current and ongoing surveys will be vital in improving our understanding of stellar evolution and binary interactions. With observations of astrophysical populations from the LVK, Gaia, Rubin, Swift, the Nancy Grace Roman Space Telescope~\citep{2013arXiv1305.5422S,2015arXiv150303757S,2019arXiv190205569A}, and more, we can place tighter empirical constraints on \ac{BPS} prescriptions and robustly quantify model uncertainties, ultimately enhancing the reliability of future astrophysical population predictions and strengthening the constraints on astrophysical processes.

\begin{acknowledgments}
The authors would like to thank Monica Gallegos-Garcia, Abhishek Chattaraj, and Jeff Andrews for useful discussions. 
AGG and DEH were supported by NSF grant PHY-2513312 and by the Simons Collaboration on Black Holes and Strong Gravity through grant SFI-MPS-BH-00012593-07.
MZ gratefully acknowledges funding from the Brinson Foundation in support of astrophysics research at the Adler Planetarium.
DM acknowledges support from the National Aeronautics and Space Administration's 2024 Graduate Research Fellowship issued through the North Carolina Space Grant.
KB acknowledges support from the Falco-DeBenedetti Career Development Professorship, \emph{Chandra} Award Numbers GO223031A, GO324037X, and GO425032X, and NASA ATP Grant 80NSSC24K0768 and LISA Preparatory Science Program Grant 80NSSC24K0361.
CR acknowledges support from NASA ATP Grant 80NSSC24K0687 to the University of North Carolina at Chapel Hill, the Research Corporation for Science Advancement (Scialog "Early Science with the LSST", award SA-LSST-2025-123a), a Charles E. Kaufman Foundation New Investigator Research Grant, the Alfred P.~Sloan Foundation, and the David and Lucile Packard Foundation.   
MZ, DM, KB, and CR acknowledge support in part from grant NSF PHY-2309135 to the Kavli Institute for Theoretical Physics (KITP), where part of this work was performed.
MMB is supported by the Swiss National Science Foundation (PI Fragos, project number CRSII5\_213497).
DEH was also supported by the NSF-Simons AI-Institute for the Sky (SkAI) via grants NSF AST-2421845 and Simons Foundation MPS-AI-00010513, and by the Kavli Institute for Cosmological Physics through an endowment from the Kavli Foundation and its founder Fred Kavli.
\end{acknowledgments}

\software{}
This work made use of the following software packages: \texttt{astropy} \citep{astropy:2013,astropy:2018,astropy:2022,astropy_14207420}, \texttt{matplotlib} \citep{Hunter:2007}, \texttt{numpy} \citep{numpy}, \texttt{pandas} \citep{mckinney-proc-scipy-2010,pandas_13819579}, \texttt{python} \citep{python}, \texttt{scipy} \citep{2020SciPy-NMeth,scipy_13352243}, \texttt{COSMIC} \citep{COSMIC,COSMIC_20721229}, \texttt{Cython} \citep{cython:2011}, \texttt{h5py} \citep{collette_python_hdf5_2014}, \texttt{POSYDON} \citep{POSv1,POSv2}, \texttt{schwimmbad} \citep{schwimmbad}, \texttt{seaborn} \citep{Waskom2021}, \texttt{METISSE} \citep{METISSE_comp,METISSE_bin,METISSE_17650929}, and \texttt{tqdm} \citep{tqdm_14047011}.

This work uses Modules for Experiments in Stellar Astrophysics, MESA \citep{Paxton2011, Paxton2013, Paxton2015, Paxton2018, Paxton2019, Jermyn2023,MESA_13353788}. The MESA EOS is a blend of the OPAL \citep{Rogers2002}, SCVH \citep{Saumon1995}, FreeEOS \citep{Irwin2004}, HELM \citep{Timmes2000}, PC \citep{Potekhin2010}, and Skye \citep{Jermyn2021} EOSes. Radiative opacities are primarily from OPAL \citep{Iglesias1993, Iglesias1996}, with low-temperature data from \citet{Ferguson2005} and the high-temperature, Compton-scattering dominated regime by \citet{Poutanen2017}. Electron conduction opacities are from \citet{Cassisi2007} and \citet{Blouin2020}. Nuclear reaction rates are from JINA REACLIB \citep{Cyburt2010}, NACRE \citep{Angulo1999} and additional tabulated weak reaction rates \citet{Fuller1985, Oda1994, Langanke2000}. Screening is included via the prescription of \citet{Chugunov2007}. Thermal neutrino loss rates are from \citet{Itoh1996}. Roche lobe radii in binary systems are computed using the fit of \citet{Eggleton1983}. Mass transfer rates in Roche lobe overflowing binary systems are determined following the prescription of \citet{Ritter1988}.

Software citation information aggregated using \texttt{\href{https://www.tomwagg.com/software-citation-station/}{The Software Citation Station}} \citep{software-citation-station-paper,software-citation-station-zenodo}.

\clearpage

\appendix

\section{Summary of stellar and binary prescriptions across codes \label{sec:pres-table}}
Section~\ref{sec:methodDetails} provides a description of the main stellar and binary evolution prescriptions used in \COS{}, \MET{}, and \POS{}. Here, we present a high-level summary of these prescriptions in Table~\ref{tab:prescriptions}. We break up the prescriptions by stellar versus binary evolution, and highlight instances where the prescriptions differ between codes in red. While we do not mention magnetic braking in Section~\ref{sec:methodDetails}, we include the prescription here since its implementation is slightly different in \POS{}.  We do not expect this difference to influence our main results appreciably. However, differences in the wind prescription and the calculation of the \ac{CE} binding energy in \COS{}, as well as the prescription for the onset of \ac{CE} in \POS{},  will lead to a large difference in how binaries evolve between codes. See Sections~\ref{sec:results} and~\ref{sec:pops} for our main findings.

\begin{table*}[htbp]
  \centering
    \begin{tabular}{|l|@{\hspace{0pt}}P{0.24\columnwidth}@{\hspace{1pt}}|@{\hspace{0pt}}P{0.24\columnwidth}@{\hspace{1pt}}|@{\hspace{0pt}}P{0.24\columnwidth}@{\hspace{1pt}}|}

    \hhline{|====|}
        \multicolumn{4}{|c|}{Stellar Evolution Prescriptions} \\

    \hhline{|====|}
        \multicolumn{1}{|c|}{Prescription} & \COS & \MET & \POS \\
        \Xhline{2\arrayrulewidth}
        Solar metallicity & 0.014 & 0.014 & 0.014 \\
        \hline
        Max NS mass & $2.5 M_\odot$ & $2.5 M_\odot$ & $2.5 M_\odot$ \\
        \hline
        Max $\nu$ mass loss & $0.5 M_\odot$ & $0.5 M_\odot$ & $0.5 M_\odot$ \\
        \hline
        Magnetic Braking & ~\cite{2003ApJ...599..516I} & ~\cite{2003ApJ...599..516I} & \cellcolor{red!25}\cite{1983ApJ...275..713R} \\ 
        \hline 
        Remnant mass & ~\cite{2012ApJ...749...91F} delayed & ~\cite{2012ApJ...749...91F} delayed & ~\cite{2012ApJ...749...91F} delayed \\
        \hline
        PISN &~\cite{Hendriks:2023yrw} & ~\cite{Hendriks:2023yrw} & ~\cite{Hendriks:2023yrw} \\
        & $\Delta M_{\rm PPI} = 0$ & $\Delta M_{\rm PPI} = 0$ & $\Delta M_{\rm PPI} = 0$ \\
        \hline
        ECSN &\cellcolor{red!25}  $m_{\rm CO} \in (1.37\mbox{--}1.43)M_\odot$ & $m_{\rm He} \in (2.6\mbox{--}2.95)M_\odot$  & $m_{\rm He} \in (2.6\mbox{--}2.95)M_\odot$ \\
        &\cellcolor{red!25}  \citep{2015MNRAS.451.2123T} &$\sim$\cite{2015MNRAS.451.2123T} &$\sim$\cite{2015MNRAS.451.2123T} \\
        \hline
        SN kicks & None & None & None \\
        \hline
         & \cellcolor{red!25} \underline{OB stars:} \cite{Vink2001} & \texttt{MESA Dutch} scheme: \underline{cool winds}:~\citet{deJager1988}, $M_{\rm ZAMS} > 8 M_\odot$; \citet{1975psae.book..229R} or \citet{Bloecker1995},  $M_{\rm ZAMS} < 8 M_\odot$&\texttt{MESA Dutch} scheme: \underline{cool winds}:~\citet{deJager1988}, $M_{\rm ZAMS} > 8 M_\odot$; \citet{1975psae.book..229R} or \citet{Bloecker1995},  $M_{\rm ZAMS} < 8 M_\odot$\\
         Winds &\cellcolor{red!25} \underline{WR stars:} \cite{Vink2005}
        & \underline{hot winds:} \citet{Vink2001}, $X_{\rm surf} > 0.4$; WR winds from~\citet{Nugis2000}, $X_{\rm surf} < 0.4$& \underline{hot winds:} \citet{Vink2001}, $X_{\rm surf} > 0.4$; WR winds from~\citet{Nugis2000}, $X_{\rm surf} < 0.4$\\
        & \cellcolor{red!25} \underline{LBV winds:} \citet{2010ApJ...714.1217B} with mass loss rate of $1.5 \times 10^{-4} M_\odot \text{yr}^{-1}$ for stars that cross HD limit & \underline{LBV winds:} \citet{2010ApJ...714.1217B} with mass loss rate of $10^{-4} M_\odot \rm yr^{-1}$ for stars that cross HD limit & \underline{LBV winds:} \citet{2010ApJ...714.1217B} with mass loss rate of $10^{-4} M_\odot \rm yr^{-1}$ for stars that cross HD limit \\
         \hline
        \hhline{|====|}
        \multicolumn{4}{|c|}{Binary Evolution Prescriptions} \\
        \hhline{|====|}
        \multicolumn{1}{|c|}{Prescription} & \COS & \MET & \POS \\
        \Xhline{2\arrayrulewidth}
        
        Tides & \cite{Hurley_BSE} & \cite{Hurley_BSE} &  \cite{Hurley_BSE} \\
        \hline
        Accretion efficiency &
        \underline{MS/HG/CHeB:} 10$\times$ accretor thermal rate  \underline{GB/EAGB/AGB:} unlimited & \underline{MS/HG/CHeB:} 10$\times$ accretor thermal rate  \underline{GB/EAGB/AGB:} unlimited & \cellcolor{red!25} \underline{Non-degenerate stars:} all mass accreted until donor critically rotating \\
        & \underline{Degenerate stars:} accretion capped at Eddington limit & \underline{Degenerate stars:} accretion capped at Eddington limit & \underline{Degenerate stars:} accretion capped at Eddington limit   \\

        \hline

        & $q_{\rm crit}$: \cite{Hurley_BSE} ; \cite{1987ApJ...318..794H} for GB/AGB stars (\texttt{qcflag = 1})
        & $q_{\rm crit}$: \cite{Hurley_BSE} ; \cite{1987ApJ...318..794H} for GB/AGB stars (\texttt{qcflag = 1}) & \cellcolor{red!25} If mass transfer rate exceeds $0.1 M_\odot \rm yr^{-1}$, if stellar radii overflows L$_2$ point, contact phase while one of stars is \\
       Onset of \ac{CE}  & If stellar radii go into contact or if stellar radii overflow components Roche radius. Periapse radius is not checked (\texttt{smt$\_$periastron$\_$check=0}) & If stellar radii go into contact or if stellar radii overflow components Roche radius. Periapse radius is not checked (\texttt{smt$\_$periastron$\_$check=0}) & \cellcolor{red!25}   post-MS, or if photon trapping radius is larger than a CO accretor's Roche radius.\\
        \hline
        \ac{CE} prescription & $\alpha-\lambda$ & $\alpha-\lambda$ & $\alpha-\lambda$ \\
        \hline
        \ac{CE} binding energy ($\lambda$) & \cellcolor{red!25} \cite{Claeys2014} & Integrates gravitational energy of envelope from exact core-envelope boundary ($X_H = 0.1$) &  Integrates gravitational energy of envelope from exact core-envelope boundary ($X_H = 0.1$)\\
        \hline
        \ac{CE} efficiency ($\alpha$) & 1 & 1 & 1 \\
        \hline
        \ac{CE} option for HG stars & optimistic & optimistic & optimistic \\
        \hline

    \end{tabular}

      \caption{Choices of stellar and binary evolution prescriptions for \COS, \MET, and \POS. See Section~\ref{sec:prescriptions} for a more detailed explanation of each prescription. Discrepancies in the prescriptions between codes are highlighted in red.}
        \label{tab:prescriptions}
  
\end{table*}

\section{Variation of common envelope parameter settings\label{sec:bin-variations}}

Although we do not explicitly consider any variations in binary physics in this work, rapid \ac{BPS} codes contain several parameters that remain poorly constrained. These settings also have no direct complement in detailed \ac{BPS} codes. Most of these parameters regulate stable and unstable mass transfer, such as the prescription used to define the onset of a \ac{CE} phase. In \COS{} and \MET{}, the onset of \ac{CE} is determined by the critical mass ratio, $q_{\rm crit} = m_{\rm donor} / m_{\rm accretor}$. Another parameter that dictates the onset of \ac{CE} is the \texttt{smt$\_$periastron$\_$check} flag, which automatically triggers a \ac{CE} event if the donor radius is greater than the periapse radius. In our main analysis, we choose \texttt{qcflag$=$1} and \texttt{smt$\_$periastron$\_$check=0} for historical consistency. However, because there is not a broad consensus on which configuration is optimal, alternative choices may be equally valid. \citet{2026arXiv260727391V} recently investigated how various \ac{CE} evolution prescriptions affect binary neutron star merger rates and found \texttt{qcflag} to have a large effect. Although \texttt{smt$\_$periastron$\_$check} was left on by default in their analysis, it is known that this flag has a substantial impact on results due to the impact the choice of \COS{} version has on final populations~\citep[earlier versions had this flag turned on by default;][]{2026arXiv260727391V}.

Here, we investigate the influence of these two flags in \COS{} and \MET{} for our fiducial $0.01 Z_\odot$ population and highlight the substantial impact on the resultant \ac{BBH} merger populations. Note that \POS{} calculates the onset of \ac{CE} self-consistently during the binary evolution, and thus all populations evolved with \POS{} will remain identical regardless of which parameters are chosen here. 

\begin{figure*}
    \centering
    \includegraphics[width=0.9\linewidth]{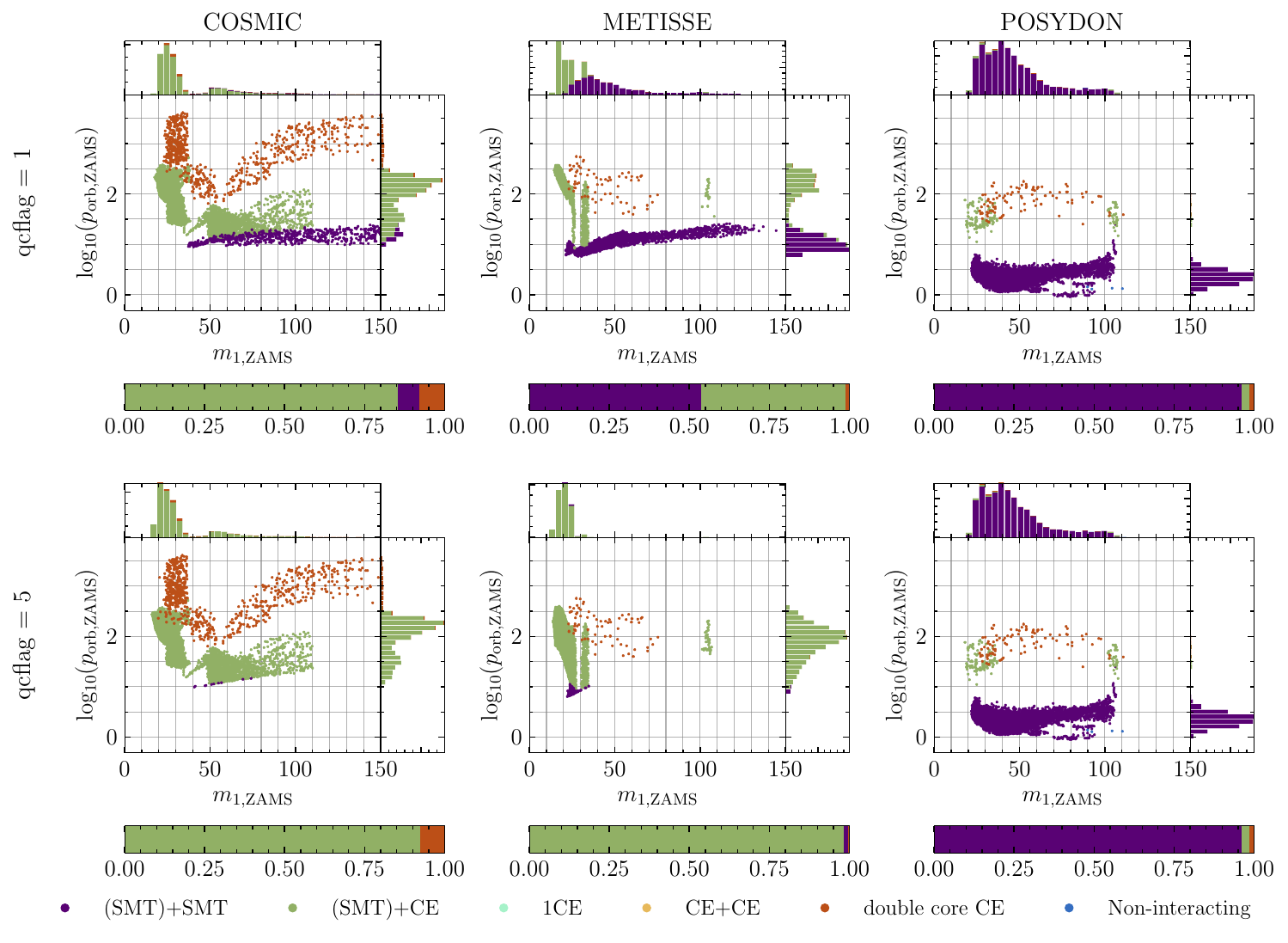}
    \caption{Same as the top panels of Figure~\ref{fig:m1-p}, but now investigating the influence of the rapid \ac{BPS} code parameter determining the critical mass ratio $q_{\rm crit}$ prescription for the onset of a \ac{CE} event. (Top row) The \ac{ZAMS} properties of BBH mergers when using \texttt{qcflag=1}. (Bottom row) The \ac{ZAMS} properties of \ac{BBH} mergers when using \texttt{qcflag=5}. Note that the \POS{} populations are identical, as \POS{} determines the onset of \ac{CE} self-consistently and does not use $q_{\rm crit}$.}
    \label{fig:qcflag}
\end{figure*}

In Figure~\ref{fig:qcflag}, we plot the progenitor properties of \ac{BBH} mergers when switching the $q_{\rm crit}$ criteria from \texttt{qcflag$=$1} to \texttt{qcflag$=$5}. \texttt{qcflag$=$5} corresponds to the $q_{\rm crit}$ values found in Section 2.3 of~\citealt{2019MNRAS.490.3740N}, and assumes mass transfer from stripped star donors is always stable. As seen in Figure~\ref{fig:qcflag}, switching \texttt{qcflag$=$1} to \texttt{qcflag$=$5} removes the majority of (SMT)$+$SMT systems in both \COS{} and \MET{}. As we found in Section~\ref{sec:scatters}, these systems typically went through Case A mass transfer. When the donor is a MS star, $q_{\rm crit} = 3$ when \texttt{qcflag$=$1}, but $q_{\rm crit}=1.717$ when \texttt{qcflag$=$5}. Since changing to \texttt{qcflag$=$5} decreases $q_{\rm crit}$, more systems will pass this threshold that otherwise would not have started a \ac{CE} event. However, since these systems start at such a low \ac{ZAMS} orbital period, they will not have enough orbital energy to eject the envelope, and instead go through a failed \ac{CE} phase. Therefore, these systems disappear altogether from the \ac{BBH} merger populations. 

\begin{figure*}
    \centering
    \includegraphics[width=0.9\linewidth]{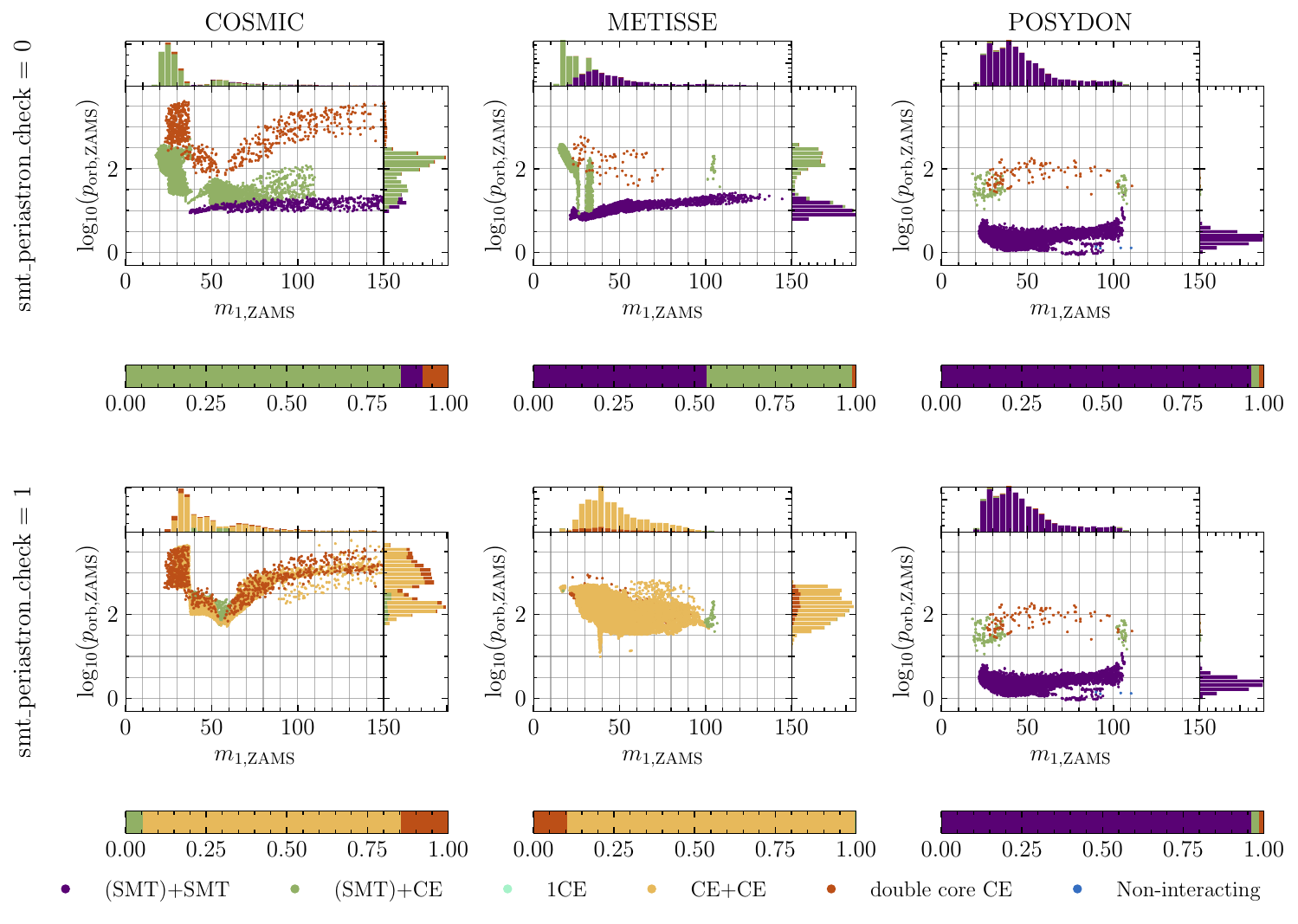}
    \caption{Same as the top panels of Figure~\ref{fig:m1-p}, but now investigating the influence of the rapid \ac{BPS} code prescription for determining the onset of \ac{CE}, \texttt{smt$\_$periastron$\_$check}. (Top row) The \ac{ZAMS} properties of BBH mergers when using \texttt{smt$\_$periastron$\_$check$=$0}. (Bottom row) The \ac{ZAMS} properties of BBH mergers when using \texttt{smt$\_$periastron$\_$check$=$1}. Note that the \POS{} populations are identical, as \POS{} determines the onset of \ac{CE} self-consistently and does not force a \ac{CE} during a contact phase.}
    \label{fig:rp-check}
\end{figure*}

In Figure~\ref{fig:rp-check}, we plot the progenitor properties of \ac{BBH} mergers when keeping \texttt{smt$\_$periastron$\_$check} off and when turning \texttt{smt$\_$periastron$\_$check} on. In \COS{} and \MET{}, orbits circularize during mass transfer phases. Therefore, when checking if the donor radius exceeds the periapse radius, \texttt{smt$\_$periastron$\_$check} will essentially check if the radii of the stars go into contact. If \texttt{smt$\_$periastron$\_$check=0}, this check is turned off, and binaries entering a contact phase may proceed through stable mass transfer. In \COS{}, the radii of stars entering the core He burning phase expand to orders of magnitude larger than they were. In \MET{}, stellar radii grow larger when on the Hertzsprung gap, but do not expand much more during the core He burning phase~\citep[see Figure 2 in][]{Maclean26}. In Section~\ref{sec:scatters}, we found that \COS{} systems with a stable mass transfer phase before the first \ac{SN} typically go through Case C mass transfer and many \MET{} systems go through Case B mass transfer. Therefore, when turning \texttt{smt$\_$periastron$\_$check} on, these systems in \COS{} and \MET{} would almost automatically enter a \ac{CE} phase since their radii are so large. This leads to most of the (SMT)$+$SMT and (SMT)$+$CE systems disappearing, and being replaced by CE$+$CE systems. The CE$+$CE systems are also required to be at larger ZAMS orbital periods to have enough orbital energy to go through two successful \ac{CE} ejections. Ultimately, these two \ac{CE} prescriptions demonstrate that the application of uncertain binary physics in rapid and hybrid \ac{BPS} codes strongly influences mass transfer scenarios, the \ac{BBH} merger efficiency, and final \ac{ZAMS} and \ac{BBH} merger demographics.

\section{\ac{BBH} merger populations for other metallicities\label{sec:otherZ}}
\begin{figure*}
    \includegraphics[width=\textwidth]{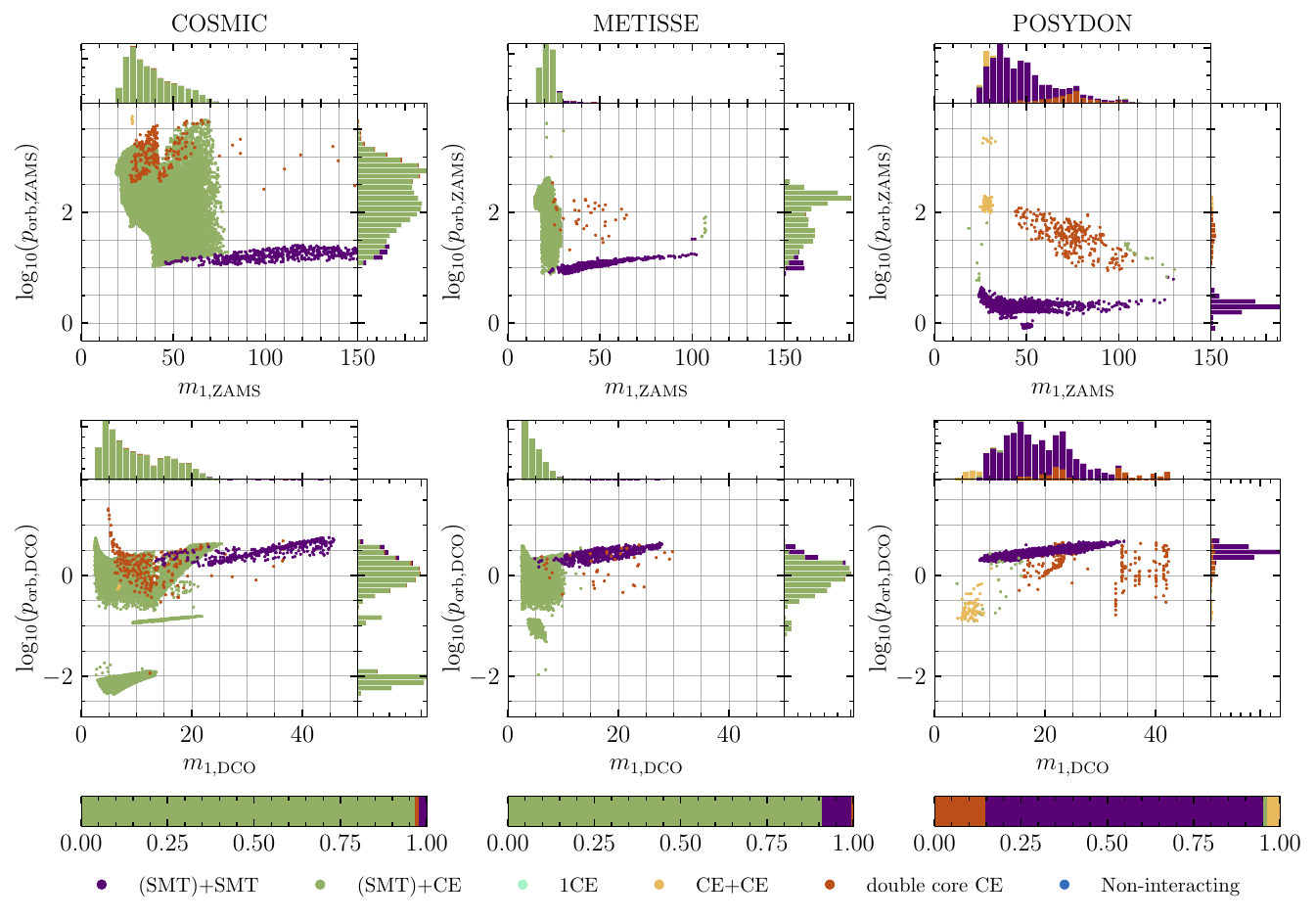}
    \caption{Same as Figure~\ref{fig:m1-p} but for our $0.1 Z_\odot$ fiducial metallicity population. }
    \label{fig:m1-p_1e-01}
\end{figure*}

\begin{figure*}
    \includegraphics[width=\textwidth]{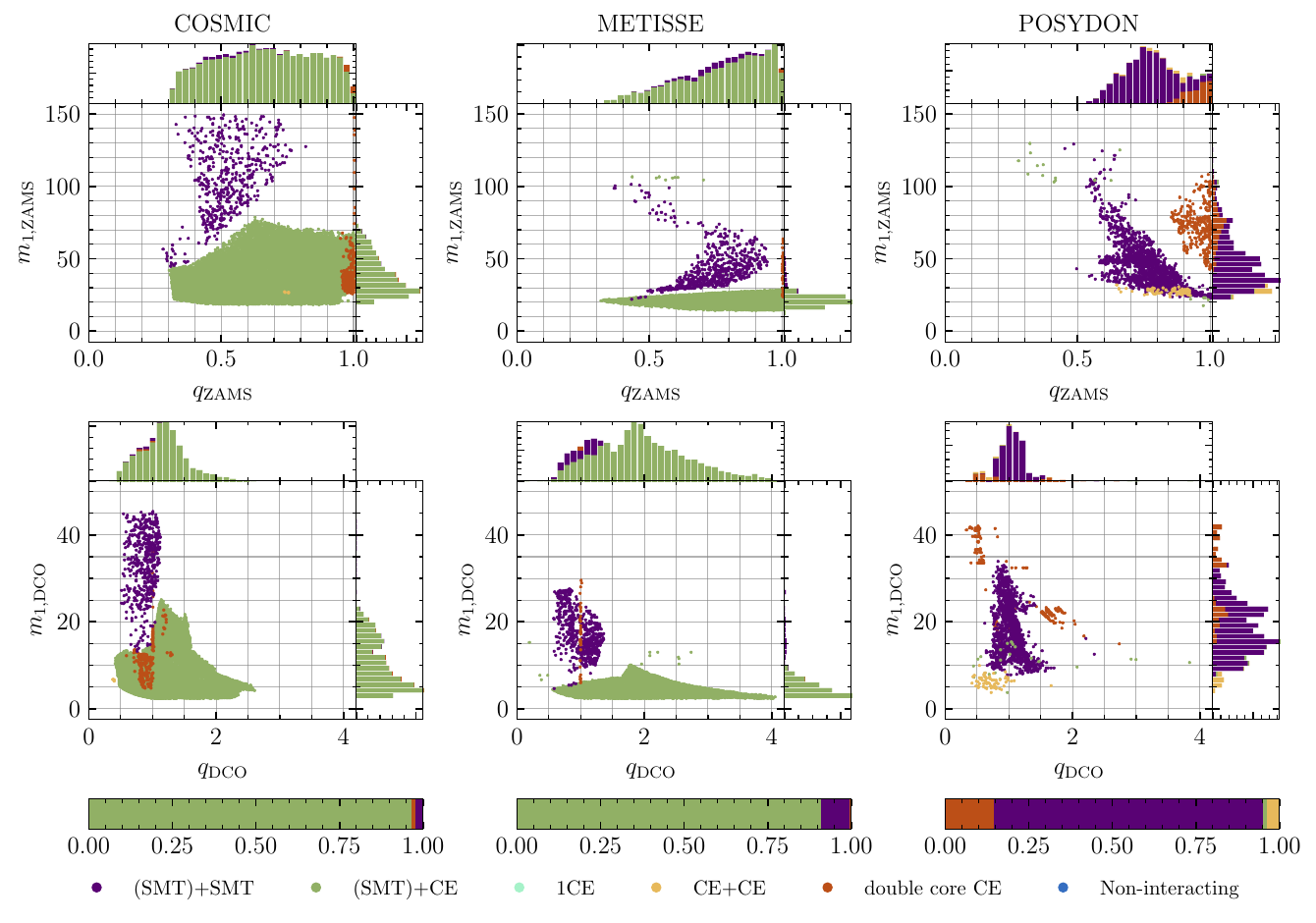}
    \caption{Same as Figure~\ref{fig:q-m1} but for our $0.1 Z_\odot$ fiducial metallicity population. }
    \label{fig:q-m1_1e-01}
\end{figure*}

\begin{figure*}
    \includegraphics[width=\textwidth]{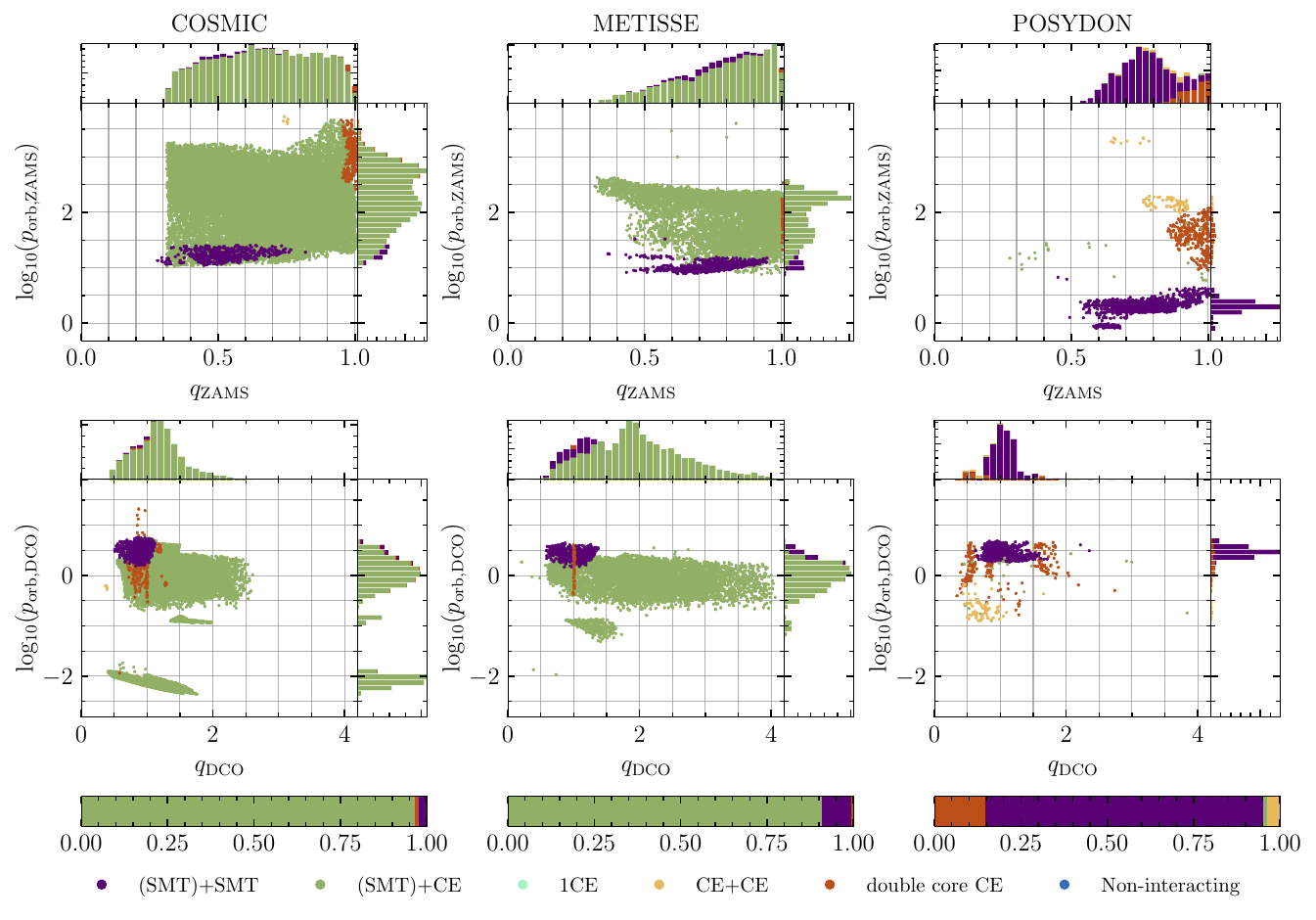}
    \caption{Same as Figure~\ref{fig:q-p} but for our $0.1 Z_\odot$ fiducial metallicity population. }
     \label{fig:q-p_1e-01}
\end{figure*}

\begin{figure*}
    \includegraphics[width=\textwidth]{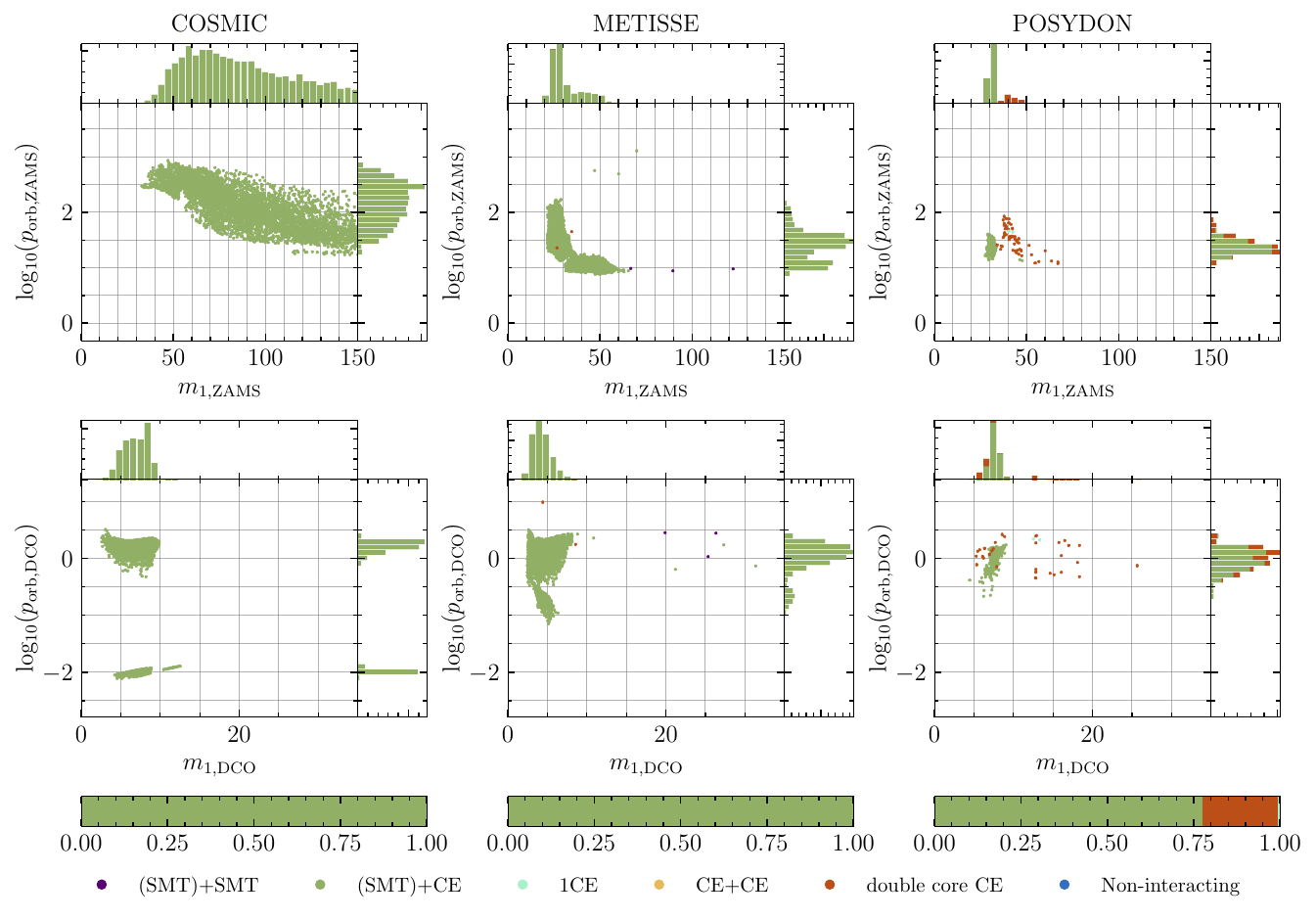}
    \caption{Same as Figure~\ref{fig:m1-p} but for our $1.0 Z_\odot$ fiducial metallicity population. }
    \label{fig:m1-p_1e+00}
\end{figure*}

\begin{figure*}
    \includegraphics[width=\textwidth]{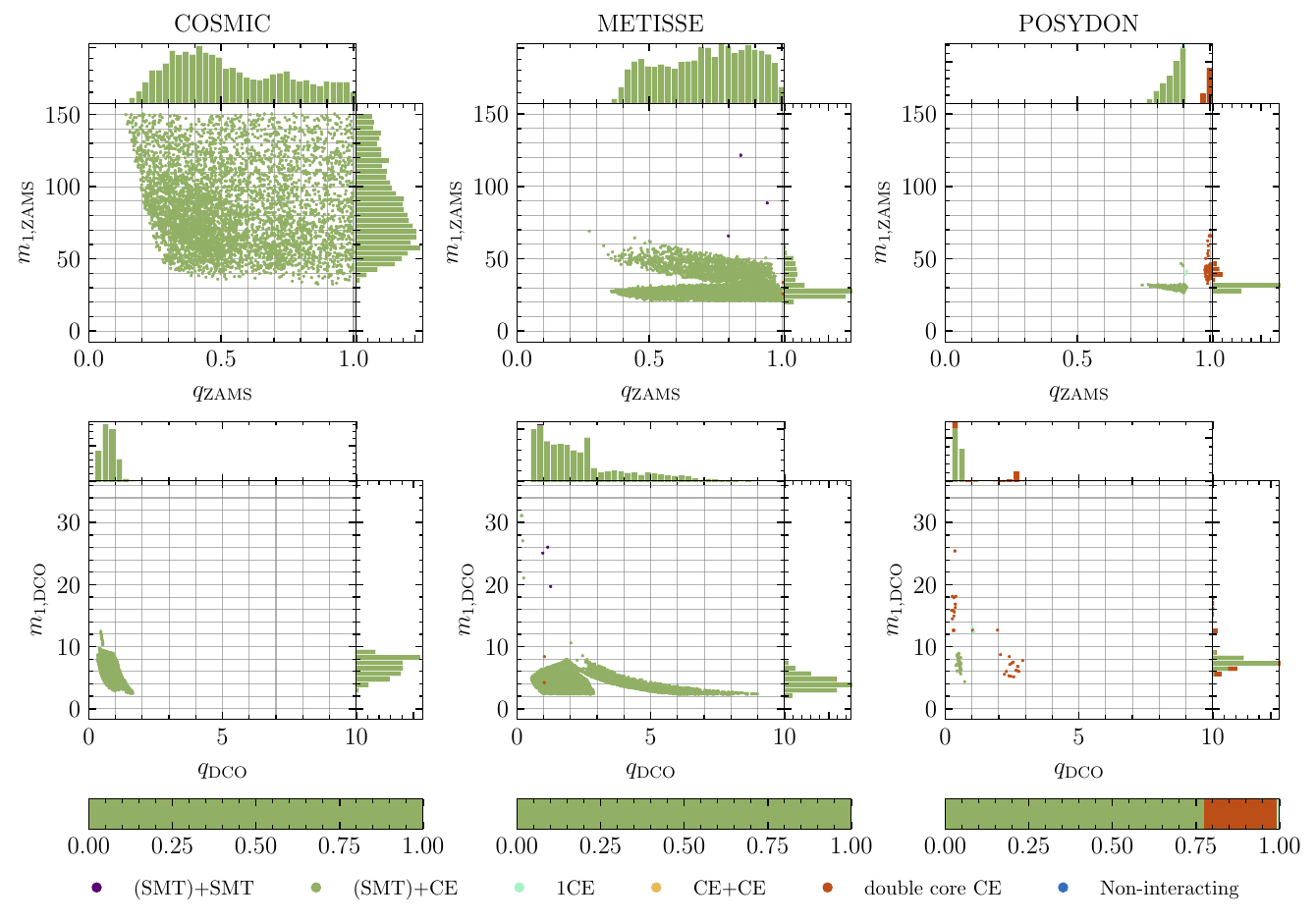}
    \caption{Same as Figure~\ref{fig:q-m1} but for our $1.0 Z_\odot$ fiducial metallicity population. }
    \label{fig:q-m1_1e+00}
\end{figure*}

\begin{figure*}
    \includegraphics[width=\textwidth]{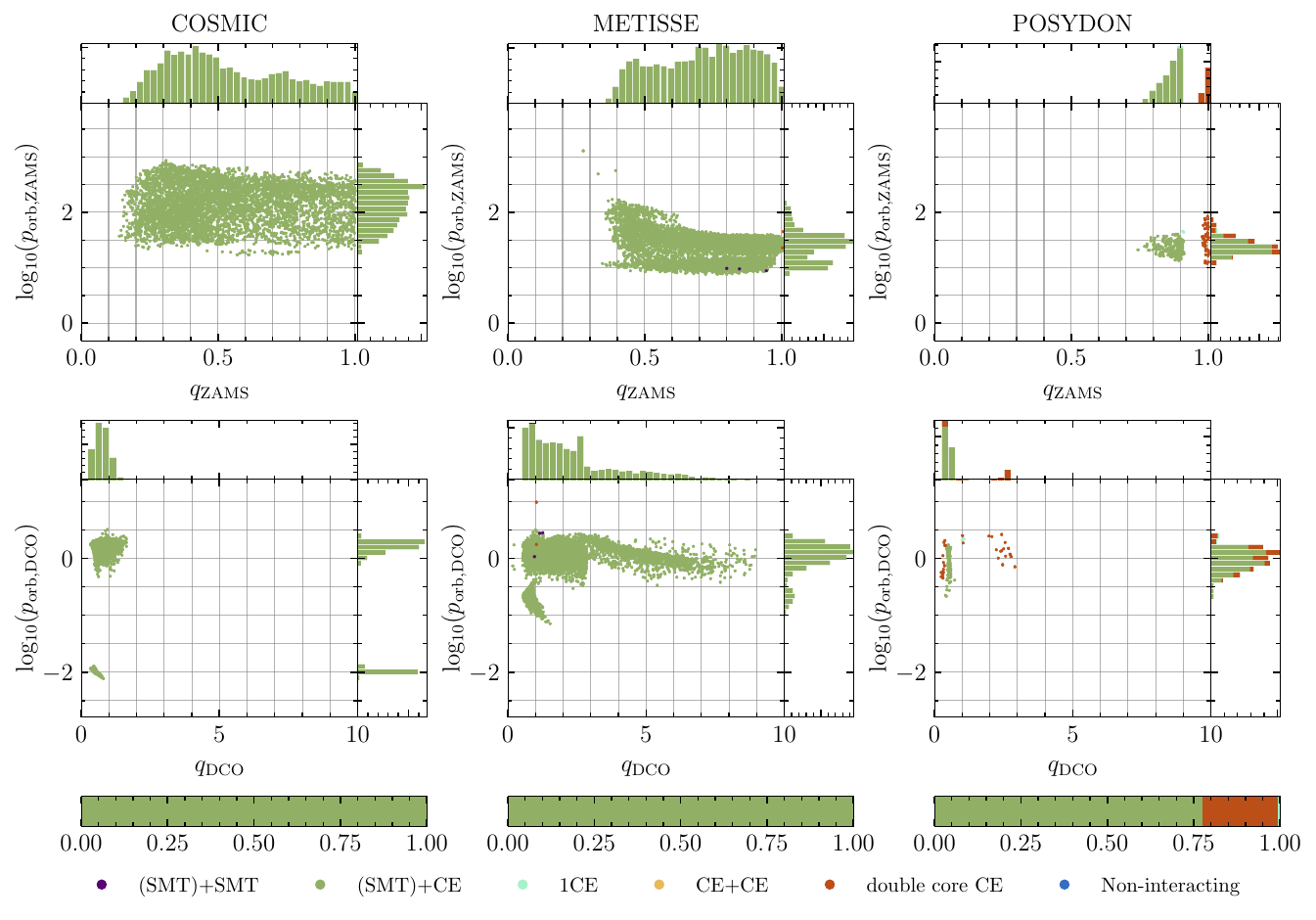}
    \caption{Same as Figure~\ref{fig:q-p} but for our $1.0 Z_\odot$ fiducial metallicity population. }
     \label{fig:q-p_1e+00}
\end{figure*}

Below, we investigate the aggregate merging \ac{BBH} populations in \COS{}, \MET{}, and \POS{} for the other two fiducial populations considered in this work, with $Z= 0.1 Z_\odot$ and $Z=1.0 Z_\odot$. In Figures~\ref{fig:m1-p_1e-01},~\ref{fig:q-m1_1e-01}, and~\ref{fig:q-p_1e-01}, we investigate the two-dimensional and marginal distributions for $(m_1,p_{\rm orb})$, $(q,m_1)$, and $(q,p_{\rm orb})$, respectively, for our $0.1 Z_\odot$ population. Figures~\ref{fig:m1-p_1e+00},~\ref{fig:q-m1_1e+00}, and~\ref{fig:q-p_1e+00} depict the equivalent distributions for the $1.0 Z_\odot$ population. The bar graphs in the bottom panels of each figure depict the percentage of \ac{BBH} mergers in each code that go through a given mass transfer scenario---CE$+$CE (gold), (SMT)$+$CE (green), (SMT)$+$SMT (purple), 1CE (cyan), double core \ac{CE} (dark orange), or non-interacting (navy). See Section~\ref{sec:scatters} for definitions of each mass transfer scenario.

As metallicity increases, there are less systems in all codes that merge in the (SMT)$+$SMT channel, with no systems merging in the (SMT)$+$SMT channel in the $1.0 Z_\odot$ populations. The \ac{BBH} merger efficiency also decreases as a function of increasing metallicity. The general correlations seen in the \ac{ZAMS} and \ac{DCO} two-dimensional distributions for each mass transfer type is similar across all three metallicity populations in each code. However, the overall \ac{ZAMS} and \ac{DCO} marginal distributions do change with metallicity. The two-dimensional and marginal distributions in \MET{} and \POS{} are robust against metallicity for each mass transfer channel. However, the overall shift to (SMT)$+$CE systems increases the relative contribution of these systems to the marginal distributions. \COS{}, on the other hand, varies more with metallicity. For example, the \ac{ZAMS} orbital period distributions shift to larger values in the $0.1 Z_\odot$ population before decreasing again in the $1.0 Z_\odot$ population. Likewise, the maximum \ac{ZAMS} primary mass in \COS{} decreases in the $0.1 Z_\odot$ population before increasing again in the $1.0 Z_\odot$ population. In both \COS{} and \MET{}, the \ac{ZAMS} mass ratio distribution extends to lower values as metallicity increases, although this effect is much larger in \COS{}, with $q_{\rm ZAMS}$ peaking around $0.4$ in the $1.0 Z_\odot$ population. While we do not go into detail on the physical reasons behind these trends, we emphasize that the metallicity dependence of various prescriptions has a large impact on the disparity between each \ac{BPS} code when predicting \ac{BBH} populations. Since realistic \ac{GW} populations convolve single-metallicity populations with a star formation history and metallicity distribution, any discrepancies in metallicity dependence across \ac{BPS} codes will compound when simulating observed \ac{GW} populations. 

\section{Population properties of merging \ac{BBH}s when including supernova kicks\label{sec:kicks}}

We now consider how the \ac{BBH} merger population changes when including \ac{SN} kicks. We evolve a new initial binary population with a fiducial metallicity of $0.01 Z_\odot$, with the following updates to \ac{SN} kicks. We use the same \ac{SN} kick prescriptions for all three codes. For core-collapse \ac{SN}, kicks are drawn from a Maxwellian distribution with a dispersion of $265~\text{km s}^{-1}$~\citep[consistent with][]{2005MNRAS.360..974H}. \ac{ECSN} are instead drawn from a Maxwellian distribution with a dispersion of $20~\text{km s}^{-1}$~\citep[consistent with][]{2019MNRAS.482.2234G}. \COS{} and \MET{} additionally consider ultra-stripped \ac{SN} for He-stars that undergo a \ac{CE} event with a compact object companion. These systems receive kicks drawn from a Maxwellian with a dispersion of $265~\text{km s}^{-1}$. Finally, \COS{} and \MET{} also consider accretion-induced collapse. These systems use the same kick distribution as for \ac{ECSN}. We use fallback-modulated kicks following~\citet{2012ApJ...749...91F}. 

\begin{figure*}
    \includegraphics[width=\textwidth]{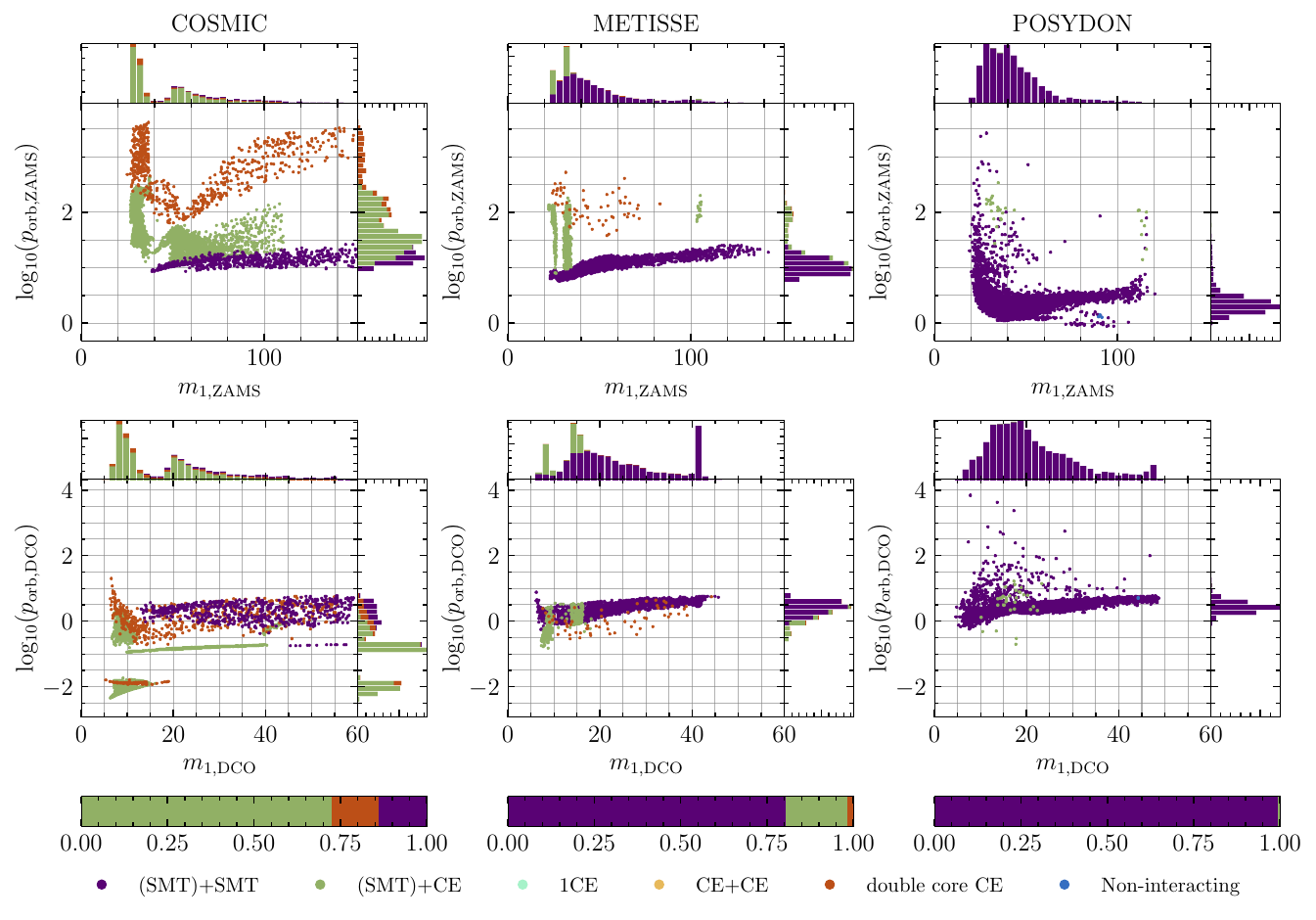}
    \caption{Same as Figure~\ref{fig:m1-p} but including \ac{SN} kicks for our fiducial $Z=0.01 Z_\odot$ population. }
    \label{fig:m1-p_kicks}
\end{figure*}

\begin{figure*}
    \includegraphics[width=\textwidth]{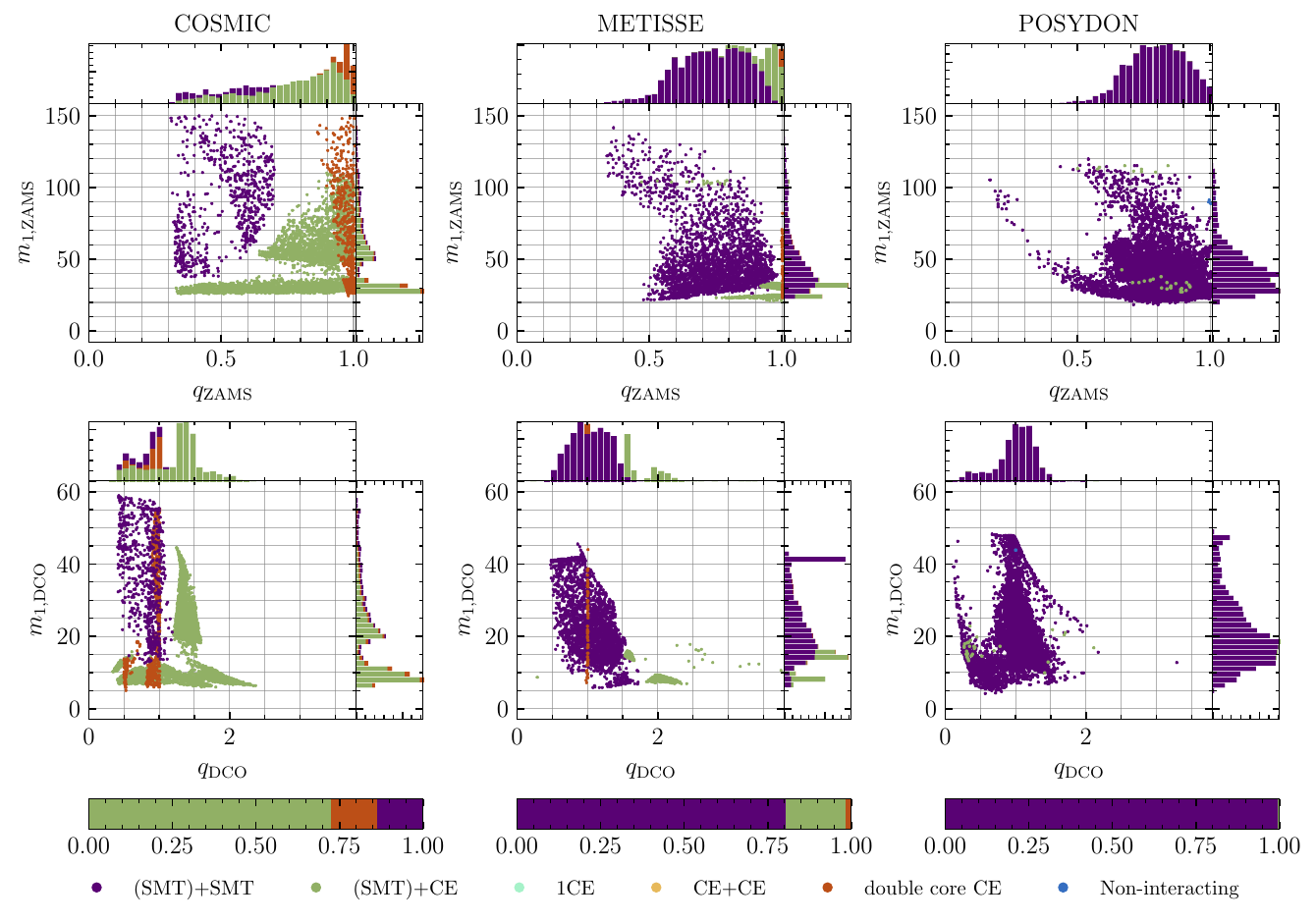}
    \caption{Same as Figure~\ref{fig:q-m1} but including \ac{SN} kicks for our fiducial $Z=0.01 Z_\odot$ population. }
    \label{fig:q-m1_kicks}
\end{figure*}

\begin{figure*}
    \includegraphics[width=\textwidth]{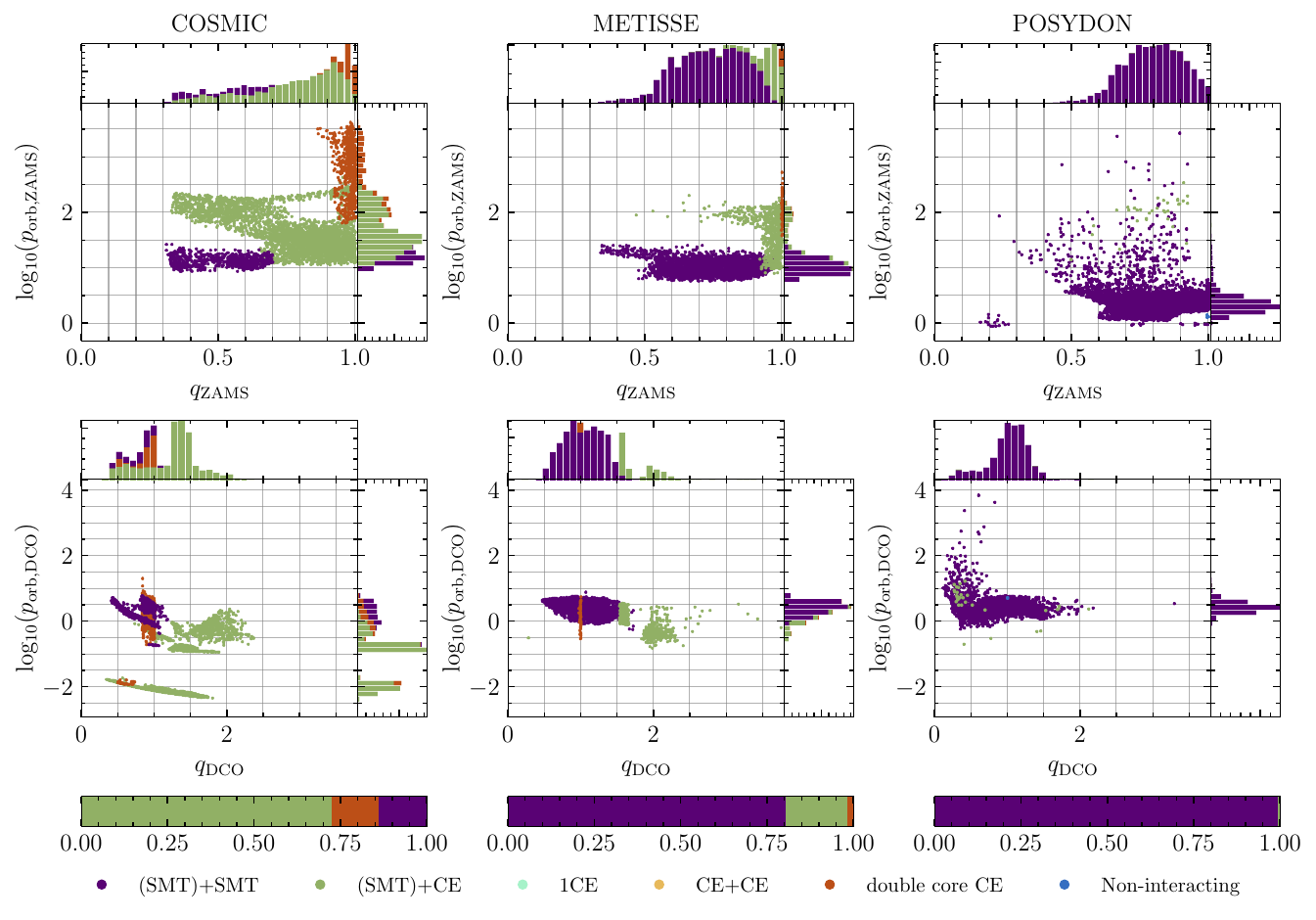}
    \caption{Same as Figure~\ref{fig:q-p} but including \ac{SN} kicks for our fiducial $Z=0.01 Z_\odot$ population. }
     \label{fig:q-p_kicks}
\end{figure*}

In Figures~\ref{fig:m1-p_kicks}, \ref{fig:q-m1_kicks}, and~\ref{fig:q-p_kicks}, we again plot the aggregate \ac{BBH} population from this new $0.01 Z_\odot$ population in the two-dimensional and marginal distributions for $(m_1,p_{\rm orb})$, $(q,m_1)$, and $(q,p_{\rm orb})$, respectively. These figures are directly comparable to Figures~\ref{fig:m1-p}, \ref{fig:q-m1}, and~\ref{fig:q-p}, but now include \ac{SN} kicks. 

\ac{BBH} progenitors evolved with \COS{} are very similar in the kick and no-kick populations, with a similar proportion of systems going through each mass transfer scenario. There are slightly more progenitors with $m_{1, \rm ZAMS}$ around $55 M_\odot$, leading to a slightly more pronounced peak near $m_{1, \rm DCO} \sim 20 M_\odot$. There are also slightly more progenitors around $p_{\rm orb, ZAMS} \sim 31$ days, and less systems that have an orbital period around 0.008 days at \ac{BBH} formation. More systems that go through the (SMT)$+$CE channel now experience mass-ratio reversal when \ac{SN} kicks are introduced.

\ac{SN} kicks seem to have a slightly less pronounced effect on \ac{BBH} progenitors in \MET{}. The \ac{ZAMS} and \ac{DCO} properties are roughly equivalent between the kicks and no-kicks populations. However, some systems that undergo the (SMT)$+$CE mass transfer scenario that originally had unequal \ac{ZAMS} mass ratios are now disrupted and no longer produce \ac{BBH} mergers when including \ac{SN} kicks, leading to a lower overall contribution of \ac{BBH} progenitors from the (SMT)$+$CE channel. 

Finally, the net population of merging \acp{BBH} in \POS{} is comparable between the kick and no-kick populations, but slightly more distinct than those in \COS{} and \MET{}. There are slightly fewer \acp{BBH} at lower \ac{ZAMS} orbital periods due to some systems getting \ac{SN} kicks that lead to a merger with the newly-formed black hole's stellar companion. However, there is now an additional population of (SMT)$+$SMT systems at higher ZAMS orbital periods at lower primary masses. These systems now merge due to the \ac{SN} kicks inducing orbital eccentricity large enough for these systems to merge in a Hubble time. There is also an additional population of unequal \ac{ZAMS} mass ratio systems due to \ac{SN} kicks saving these systems from going through failed \ac{CE} events. These results are similar to those presented in~\citet{Briel26}. These extra systems increase the proportion of total \ac{BBH} mergers sourced from the solely stable mass transfer channel. These additional (SMT)$+$SMT systems introduced in the kicks population typically have a smaller ZAMS mass ratio, do not undergo mass ratio reversal, and end up with smaller mass ratios at \ac{BBH} formation. There are a few more non-interacting systems at very low ZAMS orbital periods that likely go through chemically homogeneous evolution, survive, and merge. While we focus on the affect \ac{SN} kicks have on the $0.01 Z_\odot$ population, other trends may appear at different metallicities. Although we do not consider \ac{SN} kicks in depth in this work, the uncertainties in \ac{SN} prescriptions will add in another systematic when comparing \ac{BPS} outputs.

\bibliography{references}{}
\bibliographystyle{aasjournalv7}

\end{document}